\documentclass[powders,article,preprints,pdftex,moreauthors]{Definitions/mdpi_preprint} 

\firstpage{1} 
\pubvolume{1}
\issuenum{1}
\articlenumber{0}
\pubyear{2022}
\copyrightyear{2022}
\datereceived{} 
\dateaccepted{} 
\datepublished{} 
\hreflink{https://doi.org/} % If needed use \linebreak
\usepackage{siunitx}
\usepackage{subcaption}
\usepackage{amssymb,amsfonts}
\usepackage{dsfont}
\usepackage{bbm}
\usepackage{verbatim}
\usepackage[noEnd=true]{algpseudocodex}

\usepackage{algorithm}
\usepackage[toc,page]{appendix}

\newcommand{\SSM}{\mathrm{SSM}}
\newcommand{\MeanSSM}{\mathrm{MSSM}}
\newcommand{\GaussianFilter}{k}
\newcommand{\Z}{\mathbb{Z}}
\newcommand{\R}{\mathbb{R}}

\DeclareMathOperator*{\argmax}{arg\,max}

\newcommand{\ImageIntactParticle}{I_{\mathrm{intact}}}
\newcommand{\ImageIntact}{I_{\mathrm{intact}}}

\newcommand{\ImageBrokenParticle}{I_{\mathrm{broken}}}
\newcommand{\WindowIntactParticle}{W_{\mathrm{intact}}}

\newcommand{\WindowBrokenParticle}{W_{\mathrm{broken}}}

\newcommand{\ImagePhases}{I_{\mathrm{phases}}}
\newcommand{\NumberPhases}{n_{\mathrm{phases}}}
\newcommand{\MaskImageIntact}{M_{\mathrm{intact}}}
\newcommand{\MaskImageBroken}{M_{\mathrm{broken}}}
\newcommand{\SetMaskImageIntact}{W^{\prime}_{\mathrm{intact}}}
\newcommand{\SetMaskImageBroken}{W^{\prime}_{\mathrm{broken}}}

\newcommand{\SetFragments}{\mathcal{F}}
\newcommand{\ImagesSetFragments}{I_{\mathcal{F}}}
\newcommand{\MaskSetFragments}{M_{\mathcal{F}}}
\newcommand{\Fragment}{F}
\newcommand{\MaskFragment}{M_{\Fragment}}

\newcommand{\NumberFragments}{n_{\Fragment}}
\newcommand{\WindowFragment}{W_{\Fragment}}

\newcommand{\ImageGaps}{I^{\mathrm{gaps}}}

\newcommand{\MNCC}{\mathrm{MNCC}}

\newcommand{\FFT}{\mathrm{FFT}}
\newcommand{\fixed}{\mathrm{f}}
\newcommand{\moving}{\mathrm{m}}
\newcommand{\ImageFixed}{I^{\fixed}}
\newcommand{\ImageFixedAdjusted}{\widetilde{I}^{\fixed}}
\newcommand{\MaskFixedAdjusted}{\widetilde{M}^{\fixed}}
\newcommand{\ImageMoving}{I^{\moving}}
\newcommand{\ImageFragment}{I_\Fragment}

\newcommand{\WindowImageFixed}{W^{\fixed}}
\newcommand{\WindowImageMoving}{W^{\moving}}

\newcommand{\MaskFixed}{M^{\fixed}}
\newcommand{\MaskMoving}{M^\moving}

\newcommand{\ImageReassemblingResult}{I_{\mathrm{R}}}

\newcommand{\ImageReassemblingResultLabeled}{I_{\mathrm{R}}}
\newcommand{\ImageReassemblingResultLabeledAdjusted}{\widetilde{I}_{\mathrm{R}}}
\newcommand{\ImageFractureSurfaceReassembled}{\ImageFractureSurface_\mathrm{R}}

\newcommand{\ImageFractureSurface}{I^{\mathrm{fracture}}}
\newcommand{\FractureVoxels}{F^{\mathrm{fracture}}}

\newcommand{\VolumePhase}{\rho}

\newcommand{\MeanImageMoving}{\overline{\ImageMoving_{t,\theta}}}
\newcommand{\MeanImageFixed}{\overline{\ImageFixed_{t,\theta}}}

\newcommand{\Transformationfunction}{T_{t,\theta}}
\newcommand{\TransformationfunctionOptimal}{T_{t_{\mathrm{opt}},\theta_{\mathrm{opt}}}}
\newcommand{\ShiftFunction}{T_t}

\newcommand{\RotationFunction}{R_\theta}

\newcommand{\Tessellation}{\mathcal{T}}

\newcommand{\NumberTessellationCells}{m}
\newcommand{\FragmentGenerated}{C^{\prime}}
\newcommand{\ImageFragmentGenerated}{I_{\FragmentGenerated}}
\newcommand{\MaskFragmentGenerated}{M_{\FragmentGenerated}}
\newcommand{\WindowFragmentGenerated}{W_{\FragmentGenerated}}
\newcommand{\SetFragmentsGenerated}{\mathcal{T}^{\prime}}
\newcommand{\NumberFragmentsGenerated}{m^{\prime}}
\newcommand{\GroundTruth}{S^{\prime}}

\newcommand{\VolumeFractionsFractureSurfaces}{\rho^{\mathrm{fracture}}}
\newcommand{\RatioTransgranular}{\lambda^{\mathrm{trans}}}
\newcommand{\RatioIntergranular}{\lambda^{\mathrm{inter}}}

\newcommand{\watershed}{S}
\newcommand{\transformedImage}{I_{t,\theta}^\mathrm{m}}
\newcommand{\transformedImageOptimal}{\widetilde{I}^\mathrm{m}}
\newcommand{\transformedMaskOptimal}{\widetilde{M}^\mathrm{m}}
\newcommand{\BoundingBoxtransformedMaskOptimal}{\widetilde{D}}

\Title{Virtual reassembling of 3D fragments for the data-driven analysis of fracture mechanisms in composite materials}

\TitleCitation{Analysis of fracture mechanisms in heterogeneous structures - virtual reassembling of composite particles in 3D}

\Author{Thomas Wilhelm$^{1,}$*\orcidA{}, Trang Thu V\~{o}$^{2}$, Orkun Furat$^{1}$, Urs A. Peuker$^{2}$ and Volker Schmidt$^{1}$}

\AuthorNames{Thomas Wilhelm, Trang Thu V\~{o}, Orkun Furat, Urs A. Peuker and Volker Schmidt}

\AuthorCitation{}

\address{%
$^{1}$ \quad Institute of Stochastics, Ulm University, 89069 Ulm, Germany 
\\
$^{2}$ \quad Institute of Particle Technology and Mineral Processing, TU Bergakademie Freiberg, \mbox{09599 Freiberg, Germany}}

\corres{Correspondence: thomas.wilhelm@uni-ulm.de}

\abstract{This paper introduces a novel method for characterizing fracture mechanisms in composite materials using 3D image data gained by computed tomography (CT) measurements. In mineral liberation, the understanding of these mechanisms is crucial, particularly whether fractures occur along the boundaries of mineral phases (intergranular fracture) and/or within mineral phases (transgranular fracture). Conventional techniques for analyzing fracture mechanisms are focused on globally comparing the surface exposure of mineral phases extracted from image measurements before and after fracture. Instead, we present a virtual reassembling algorithm based on image registration techniques, which is applied to 3D data of composite materials before and after fracture in order to determine and characterize the individual fracture surfaces. This enables us to conduct a local quantitative analysis of fracture mechanisms by voxelwise comparing adjacent regions at fracture surfaces. A quantitative analysis of fracture mechanisms is especially important in the context of geometallurgical recycling processes. As primary deposits are decreasing worldwide, the focus is shifting to secondary raw materials containing low concentrations of valuable elements such as lithium. To extract these elements, they can be enriched as engineered artificial minerals in the slag phase of appropriately designed cooling processes. The subsequent liberation through comminution processes, such as crushing, is essential for the extraction of valuable minerals. A better understanding of crushing processes, especially fracture mechanisms in slags, is crucial for the success of recycling. The reassembling algorithm presented in this paper is evaluated through a simulation study, followed by an application to a naturally occurring ore and a slag resulting from a recycling process. 
}


\begin{document}

\section{Introduction}\label{Sec:Introduction}

Fracture mechanisms causing mineral liberation was initially investigated by Gaudin~\cite{Gaudin1939}, who differentiated between fracture along the boundaries of mineral phases (intergranular fracture) and random fracture (transgranular fracture). The latter is often considered to be the predominant mechanism for liberation as it is not influenced by mineral characteristics, like hardness differences of various minerals phases~\cite{Hsih1994}, nor by mineral morphology~\cite{Singh2014}. However, in reality, pure transgranular fracture does not occur, but fracture occurs as a combination of transgranular and intergranular fracture~\cite{KingSchneider1998}. Recently, Mariano et al.~\cite{Mariano2016} provided a summary of the definitions of random fracture used by various authors. In addition to investigating the complex interplay of various factors that influence fracture, several methods are proposed in the literature for quantifying fracture mechanisms in comminution processes. For example, Little et al.~\cite{Little2016} demonstrated the presence of intergranular fracture in ores by analyzing the preservation of the shape of mineral phases and comparing the degree of liberation. More recently, Lei{\ss}ner et al.~\cite{Leissner2016}, Mirzaei and Khalesi~\cite{MirzaeiKhalesi2019} have introduced quantitative approaches to determine the fraction of intergranular and transgranular fracture based on 2D analysis of mineral surface exposure. However, 3D data are essential for adequate quantification of fracture mechanisms. A first step in this direction was taken in~\cite{Xu2013}, where fracture mechanisms were investigated using 3D measurements, focusing on the analysis of interfacial areas between copper mineral particles and host rock, computing the specific interfacial fraction before and after fracture. Nevertheless, these studies primarily focus on the global characterization of fracture mechanisms by comparing phase-based descriptors such as surface exposure or interfacial area before and after fracture.

\begin{figure}[ht!]
    \centering
    \includegraphics[scale=1.5]{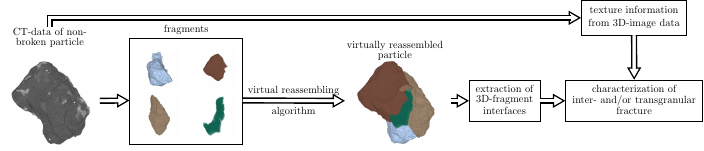}
    \caption{Workflow of the virtual reassembling of 3D fragments for the data-driven analysis of fracture mechanisms in composite materials. Initially, various fragments are reassembled to determine the interfaces between the fragments (fracture surfaces). These surfaces, along with the texture information  from 3D  measurements of the non-broken particle, are then utilized to locally  analyze the presence of intergranular or transgranular fracture.%This  provides valuable insight into the breaking mechanism occurring within the corresponding mineral.
    }
    \label{Fig:Schematic workflow analysis of breaking mechanism}
\end{figure}

In the present paper, we propose a virtual reassembling algorithm based on image registration techniques utilizing 3D CT measurements of composite materials before and after fracture. More precisely, before fracture a single non-broken particle is visualized in a CT image and, after fracture, another CT image shows how the particle has been broken into smaller fragments. By reassembling the non-broken particle from fragments, we can voxelwise extract the fracture surfaces. This detailed information about the fracture surfaces is essential for the quantitative characterization of fracture mechanisms. In particular, this method can be used to locally determine whether fractures occur predominantly intergranular, transgranular or as a superposition of these fracture mechanisms by comparing adjacent regions at fracture surfaces. In more detail, in the reassembling algorithm we employ an image registration technique that involves translation and rotation of a given pattern (so-called moving image) within a larger image (so-called fixed image)~\cite{Brown1992}. In our case, the moving image corresponds to a 3D image of a fragment, whereas the fixed image represents the non-broken particle. Using the reassembling algorithm, we iteratively reassemble the individual fragments, where we consider a specific order on how to reassemble the fragments. After reassembling it is possible to extract the fracture surfaces as regions, where two reassembled fragments are adjacent and to transfer these fracture surfaces to the non-broken particle. By characterizing the 3D microstructure of the non-broken particle in this way, i.e., knowing the positions of fracture surfaces within the non-broken particle, we are able to voxelwise specify if there is an intergranular or transgranular fracture. The workflow of this kind of a local quantitative analysis of breaking mechanisms is sketched in Figure~\ref{Fig:Schematic workflow analysis of breaking mechanism}. 

\medskip

In the literature, there exist various approaches for image registration, with some involving the placement of markers on both (moving and fixed) images, followed by an alignment of the markers~\cite{Maintz1998}. Other methods include matching extracted features or surfaces~\cite{Hill2001}. Recently, much attention has been directed towards intensity-based approaches, where intensity values are utilized to compute similarity measures between two images~\cite{Barnea1972}. Intensity-based registration typically does not require extensive preprocessing, such as segmentation or feature extraction. In the present paper, we employ an iterative 3D intensity-based approach for the registration of a fragment's image, considering six degrees of freedom. This involves finding the translations in three directions and rotations around the three axes for each fragment, where the translation of a single fragment is determined by means of a mask-based image registration approach, which is implemented via fast Fourier transformation (FFT)~\cite{Padfield2012}. Thus, this approach involves so-called masking, which is essential in the registration of fragments. It ensures that regions in the image of the non-broken particle and regions in the image of the fragments, which can wrongly influence the registration result, are ignored when computing the similarity measure to evaluate the registration result. Masked images are used to consider only those regions in the images where the non-broken particle body and the fragments are present, i.e., the background is ignored in both images. Optimal rotations are determined by means of global particle swarm optimization (PSO)\cite{Kennedy1995, Kennedy2006}. This technique has proven successful in biomedical image registration, as demonstrated by Wachowiak et al.~\cite{Wachowiak2004}. Furthermore, for evaluating how closely intensity values of the moving image match the intensity values of the fixed image, we consider a normalized cross-correlation coefficient as similarity measure. Registration in the Fourier domain using normalized cross-correlation is well-suited due to its robustness and short computation time, which has been extensively studied in the literature~\cite{Bracewell1965, Anuta1970, Padfield2012, Barnea1972, Lesse1971, Lewis1995}.

In the present paper, we first evaluate the goodness of fit of the reassembling algorithm by means of a simulation study. Here, we utilize CT data of a non-broken particle and virtually generate fragments using distance-based tessellation models, which are commonly employed for creating random grain architectures of polycrystalline materials~\cite{Okabe2000, Sedivy2016, Petrich2021}. We adapt these models to generate a grain architecture within the non-broken particle by employing the Euclidean distance as the tessellation distance, which results in a simplified tessellation also referred to as a Voronoi diagram or Voronoi tessellation. The cells of the Voronoi tessellation define a grain architecture, where a subset of the resulting cells is used as generated fragments. This fragment generation allows us to know the original positions of the fragments within the non-broken particle, serving as ground truth for evaluating the reassembling algorithm. After applying this algorithm to the fragments obtained from the Voronoi tessellation, we compare the algorithm's output with the ground truth. This comparative analysis enables us to assess the accuracy and reliability of the algorithm in reconstructing the fragments and, thus, obtaining results consistent with the original tessellation. 

\begin{figure}[ht!]
    \centering
      \includegraphics[width=0.9\textwidth]{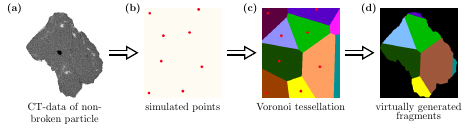}
    \caption{Workflow of virtual fragment generation. Initially, a point pattern (b) is generated within the sampling window of the non-broken particle (a). Subsequently, a Voronoi tessellation is constructed based on this point pattern, resulting in a decomposition of the sampling window into cells (c). A subset of these cells serve as virtually generated fragments (d).}
    \label{Fig:Schematic illustration of virtual fragment generation}
\end{figure}

A quantitative analysis of fracture mechanisms is especially important in the context of geometallurgical recycling processes. In view of the progressive depletion of primary deposits on the world market, there is growing interest in the development of resources from secondary raw materials. However, they often contain low concentrations of valuable elements, with lithium being a prime example. Nevertheless, the increased attention on secondary raw materials stems from the need to explore alternative sources and implement sustainable practices to meet the increasing demand for valuable elements in various industries. This shift is a strategic response to the challenges posed by the decreasing availability of primary deposits and the need to secure a more sustainable resource base. A well-known approach to extract valuable elements from secondary raw materials involves enriching them as engineered artificial minerals in the slag phase of geometallurgical recycling processes. %through an appropriately designed cooling process. 
A comprehensive understanding of generation and comminution processes of slags for the optimization of subsequent separation processes, aimed at enriching valuable minerals, requires insight into the fracture mechanisms within the slag. Moreover, the understanding of fracture mechanisms in naturally occurring composite materials is also of great interest, particularly in the mining industry, since also here the aim is to extract valuable materials, e.g. from ores. Therefore, we apply the reassembly algorithm, proposed in the present paper, to two different kinds of composites, where we characterize the fracture mechanisms of a greisen ore as well as of a slag produced in a real geometallurgical recycling process.

The rest of this paper is organized as follows. In Section~\ref{Sec:Material and sample preparation}, we provide a description of the particulate composite materials considered in this paper and how these materials are fractured by in-situ experiments. Section~\ref{Sec:Image processing} outlines the segmentation methods employed, including the separation of foreground (particle phase) from the background by means of binarization. Subsequently, a phase-based segmentation is performed to extract and analyze different mineral phases observed in CT image data. Section~\ref{Sec:Virtual reassembling algorithm} presents the reassembling algorithm, which is evaluated in a simulation study by means of goodness-of-fit measures, see Section~\ref{Sec:Simulation study}. Then, in Section~\ref{Sec:Fracture characterization}, the reassembling algorithm is applied to two different composite materials, where their fracture mechanisms are analyzed through a local quantitative fracture characterization. Section~\ref{Sec:Conclusion} concludes.

%%%%%%%%%%%%%%%%%%%%%%%%%%%%%%%%%%%%%%%%%%%%%%%%%%%%%%%%%%%%%%%%%%%%%%%%%%%%%%%%%%%%
%%%%%%%%%%%%%%%%%%%%%%%%%%%%%%%%%%%%%%%%%%%%%%%%%%%%%%%%%%%%%%%%%%%%%%%%%%%%%%%%%%%%

\section{Materials and sample preparation}\label{Sec:Material and sample preparation}

In this paper, a greisen ore from the Zinnwald deposit in the Erzgebirge region is considered for characterizing the fracture mechanisms in natural occurring composite materials. The main component in this ore is quartz, followed by zinnwaldite, topaz, muscovite, and traces of further minerals. The greisen ore was crushed using a jaw crusher and a cone crusher, and subsequently, it was further reduced to particles with sizes of approximately $\SI{4}{\milli\meter}$ using a flat cone crusher, followed by sieving and dividing. A particle with a size from the $4-\SI{5}{\milli\meter}$ size fraction was selected for in-situ compression, described below. Notably, the greisen ore does not exhibit dendritic structures.

Additionally, a lithium-aluminum slag was chosen for in-situ compression experiments, especially for its unique manufacturing and cooling process conducted at the IME Process Metallurgy and Metal Recycling - RWTH Aachen. At first, a selection of salts has been heated to their melting temperature, allowing for the homogenization of the melt. Then, the melt underwent controlled cooling at a rate of $\SI{50}{\kelvin\per\hour}$. The resulting slag exhibits a dendritic microstructure which indicates that the cooling rate was too large
for a controlled crystallization and enrichment of specific minerals. Thus, the microstructure of the slag differs significantly from the naturally occurring greisen ore, which results from slower cooling rates. Analogously to the greisen-ore particle, a slag particle with a size of $4-\SI{5}{\milli\meter}$ was selected for in-situ compression.

In order to acquire 3D image data of the non-broken greisen/slag particles and the fragments after fracture, a load cell was coupled with a CT measuring device. More precisely, the CT5000 $\SI{5}{\kilo\newton}$ in-situ load cell from Deben UK, customized for the Zeiss Xradia Versa 510 CT system, was utilized. The non-broken particle is placed in the load cell between ceramic pistons that are held in position by a guiding tube and measured at $\SI{5}{\newton}$, a force selected to ensure particle stability and prevent movement during the CT measurement. The particle is gradually loaded until a drop in force in the force-displacement curve and a change in X-ray transmission can be seen in the live projection image, indicating that the particle is broken. Following the force drop, the stress on the particle will be stopped only when the force increases again to ensure that the crack remains opened. Subsequently, another CT measurement is conducted. The measurement and reconstruction parameters for an in-situ measurement series, corresponding to a either the ore or slag particle, have been chosen consistently, see Table~\ref{Tab:Measurement settings}.

\begin{table}[ht!]
\caption{CT measurement and reconstruction settings for in-situ compression of ore and slag particle}
    \centering
    \begin{tabular}{||>{\centering\arraybackslash}p{5cm}|>{\centering\arraybackslash}p{2cm}|>{\centering\arraybackslash}p{2cm}|>{\centering\arraybackslash}p{3cm}|>{\centering\arraybackslash}p{3cm}||}
    \toprule
         &  \multicolumn{2}{c|}{ore} & \multicolumn{2}{c||}{slag}\\
         \midrule
         \textbf{measurement settings}& non-broken & broken & non-broken & broken\\ 
         \midrule
         source distance in $\SI{}{\milli\meter}$& $70$&$70$&$70$&$70$\\
         detector distance in $\SI{}{\milli\meter}$& $200$&$200$&$200$&$200$  \\
         optical magnification&  $0.4$ x & $0.4$ x& $0.4$ x& $0.4$ x \\
         acceleration voltage in $\SI{}{\kilo\volt}$&$80$&$80$&$80$&$80$   \\
         electrical power in $\SI{}{\watt}$&$7$&$7$&$7$&$7$ \\
         source filter (Zeiss standard)& LE4&LE4&LE4&LE4 \\
         voxel size in $\SI{}{\micro\meter}$&$17.8$&$17.8$&$17.8$&$17.8$ \\
         camera binning& $2$&$2$&$2$&$2$  \\
         number of projections& $401$ & $401$ & $1601$ & $1601$ \\
         exposure time in $\SI{}{\second}$& $5$&$5$&$5$&$5$ \\
         angle range in $^\circ$& $360$&$360$&$360$&$360$ \\
         \midrule
         \textbf{reconstruction settings}& \multicolumn{1}{c|}{}& \multicolumn{1}{c|}{}& \multicolumn{1}{c|}{} & \multicolumn{1}{c||}{}  \\
         reconstruction algorithm& FBP&FBP&FBP&FBP \\
         center shift&  $-0.56$ & $-0.48$ & $-0.866$ & $-0.804$\\
         defect correction&  none&none&none&none\\
         byte scaling& $(-0.02,0.057)$ & $(-0.015,0.06)$ & $(-0.003670,0.0411)$ & $(-0.003670,0.0411)$ \\
         beam hardening constant& $0.05$ & $0.05$ & $0.02$ & $0.02$ \\
         \bottomrule
    \end{tabular}
    \label{Tab:Measurement settings}
\end{table}

%%%%%%%%%%%%%%%%%%%%%%%%%%%%%%%%%%%%%%%%%%%%%%%%%%%%%%%%%%%%%%%%%%%%%%%%%%%%%%%%%%%%

\section{Image processing}\label{Sec:Image processing}

A quantitative characterization of the 3D microstructure of a particle before fracture and, thus, the characterization of the fracture mechanisms cannot be performed directly on image data, which is obtained from CT measurements as described in Section~\ref{Sec:Material and sample preparation}. Some image preprocessing is required for further characterization. First, we apply a binarization step to the CT image data of the particle before and after fracture to separate the foreground (particle) from the background in the CT images (Section~\ref{Sec:Particle-based segmentation}). This is followed by a phase-based segmentation step to extract mineral phases from the non-broken particle (Section~\ref{Sec:Phase-based segmentation}). Then we apply a watershed-based segmentation step to the image data of the particles after fracture to obtain an image of each fragment (Section~\ref{Sec:Fragment-based segmentation}), as the fragments are generally connected with each other in the raw image data of the broken particle, i.e., they are present as one contiguous region.

For each composite particle we obtained grayscale images by CT measurements before and after fracture, which are formally defined as mappings $I_{\mathrm{intact}}:\WindowIntactParticle\rightarrow\{0,\dots,65535\}$ and $\ImageBrokenParticle:\WindowBrokenParticle\rightarrow\{0,\dots,65535\}$, where the sets $\WindowIntactParticle$ and $\WindowBrokenParticle$ of voxels are finite subsets of $\Z^3$. The mapping $I_{\mathrm{intact}}$ assigns each voxel $x\in \WindowIntactParticle$ to its respective grayscale value $I_{\mathrm{intact}}(x)$. Analogously, $\ImageBrokenParticle$ assigns each voxel $x\in \WindowBrokenParticle$ to its respective grayscale value $\ImageBrokenParticle(x)$. 

%%%%%%%%%%%%%%%%%%%%%%%%%%%%%%%%%%%%%%%%%%%%%%%%%%%%%%%%%%%%%%%%%%%%%%%%%%%%%%%%%%%%

\subsection{Particle-based segmentation}~\label{Sec:Particle-based segmentation}
To separate the foreground (particle) from the background in the CT images before and after fracture, we employ a multi-step segmentation process. Initially, we reduce noise in the CT images by applying a Gaussian filter with a standard deviation of $0.2$. Subsequently, an unsharp mask filter~\cite{Pratt2007} is applied to enhance edges in the image. Then, each image is binarized using the ISODATA thresholding method~\cite{Ridler1978,Szegin2004} to distinguish foreground from background. After binarization, we obtain an image $\MaskImageIntact:\WindowIntactParticle\rightarrow\{0,1\}$ for the non-broken particle, which assigns each voxel either to the foreground or background of the particle, i.e., $\MaskImageIntact$ is given by
\begin{align}
   \MaskImageIntact(x) = \begin{cases}
       1,  \qquad &\text{if } x \text{ belongs to the foreground},\\
       0, \qquad &\text{otherwise},
   \end{cases} 
\end{align}
for each $x\in\WindowIntactParticle$. Analogously, we obtain the image $\MaskImageBroken:\WindowBrokenParticle\rightarrow\{0,1\}$, which assigns each voxel either to the foreground or background of the broken particle. Furthermore, denote the sets of voxels, which are assigned to the foreground of the non-broken and broken particle by $\SetMaskImageIntact=\{x\in\WindowIntactParticle:\MaskImageIntact(x)=1\}$ and $\SetMaskImageBroken=\{x\in\WindowBrokenParticle:\MaskImageBroken(x)=1\}$, respectively. Furthermore after binarization, we set $\ImageIntactParticle(x)=0$ for all voxels $x\in\WindowIntactParticle$ for which $\MaskImageIntact(x)=0$. Analogously, we set $\ImageBrokenParticle(x)=0$ for all voxels $x\in\WindowBrokenParticle$ for which $\MaskImageBroken(x)=0$.

This allows us to check if the extracted non-broken particle and the fragments of the broken particle have the same volume. Note that a large discrepancy between these volumes would indicate an insufficient preprocessing result or missing fragments. Therefore, we computed the volume
of the extracted non-broken particle as well as the volume of the fragments of the broken particle, and checked if $\lvert\SetMaskImageBroken\rvert/\lvert\SetMaskImageIntact\rvert\approx 1$, where $\lvert\cdot\rvert$ denotes cardinality. It turned out that $\lvert\SetMaskImageBroken\rvert/\lvert\SetMaskImageIntact\rvert=0.993$ for the ore particle, and 
 $\lvert\SetMaskImageBroken\rvert/\lvert\SetMaskImageIntact\rvert=0.995$ for the slag particle, which means that the image measurement and processing steps discussed so far seem to be correct. Minor disparities in particle volumes before and after fracture may arise from partial volume effects and the potential absence of small fragments that are not captured in the CT images after fracture. Figures~\ref{Fig:Visualization particles}a and ~\ref{Fig:Visualization particles}c show 2D slices of the CT images before fracture of both composite particles after binarization.

\subsection{Phase-based segmentation}~\label{Sec:Phase-based segmentation}
This preprocessing step enables the extraction of different mineral phases, where we assume that they can be identified by means of the grayscale values observed
 in the CT images, facilitating a quantitative analysis of the phase-based fracture mechanisms. To extract different mineral phases from the image measurements of the particles before fracture, we employ a $k$-means clustering algorithm~\cite{Dhanachandra2015}. Here, for each $x\in\WindowIntactParticle$ the grayscale values of the image $\ImageIntactParticle$ in a $3\times 3\times 3$ neighborhood are considered as features for identifying $k>1$ clusters using the $k$-means algorithm. This clustering technique assigns each foreground voxel its corresponding label from the set $\{1,\dots,k\}$, based on its feature vector, i.e., the grayscale values in the voxel's $3\times 3\times 3$ neighborhood~\cite{Gan2020}. Each label corresponds to a distinct mineral phase present in the non-broken particle. The number $k$ of clusters for each dataset is manually chosen based on visual inspection. 
 
 In addition to the mineral phases within the slag body, there are also air bubbles present. Initially identified as part of the background, these air bubbles are subsequently labeled as an additionally phase in the phase-based segmentation. Note that due to partial volume effects, the grayscale values of voxels located close to the background are influenced by neighboring particles or the background. Thus, this makes it challenging to correctly label mineral phases especially at the boundary of particles. Nevertheless, in the present paper we provide a phase-based segmentation for identifying mineral phases, which can be further improved incorporating other measurement techniques like diffraction contrast tomography or correlative tomography~\cite{Maire2014}.
%, which can identify mineral phases. 

\begin{figure}[ht!]
     \centering
     \begin{subfigure}[]{0.15\textwidth}
         \centering
         \includegraphics[width=\textwidth]{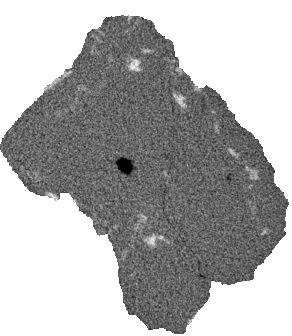}
             \subcaption{~}
     \end{subfigure}
     \hspace{1cm}
     \begin{subfigure}[]{0.15\textwidth}
         \centering
         \includegraphics[width=\textwidth]{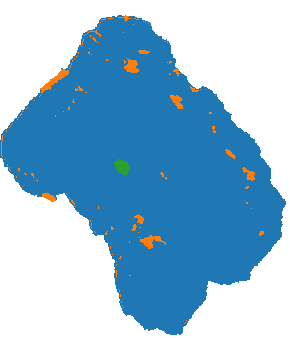}
             \subcaption{~}
     \end{subfigure}
     \hspace{1cm}
     \begin{subfigure}[]{0.14\textwidth}
         \centering
         \includegraphics[width=\textwidth]{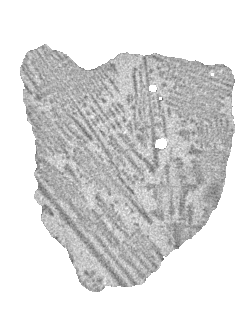}
             \subcaption{~}
     \end{subfigure}
     \hspace{1cm}
     \begin{subfigure}[]{0.14\textwidth}
         \centering
         \includegraphics[width=\textwidth]{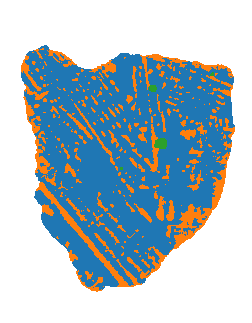}
             \subcaption{~}
     \end{subfigure}
        \caption{2D slices of CT images depicting composite particles before fracture (a: ore, c: slag) and after binarization with various mineral phases indicated by colors for the ore (b) and slag particle (d). Three different mineral phases have been identified for both particles, where phase $P_1$ is colored in blue, phase $P_2$ in orange and phase $P_3$ in green.}
        \label{Fig:Visualization particles}
\end{figure}

After extracting $\NumberPhases\geq 1$ different mineral phases, we obtain a phase-based segmentation $\ImagePhases:\WindowIntactParticle\rightarrow\{0,1,\dots,\NumberPhases\}$ of $\ImageIntact$, which assigns each voxel a mineral phase or the background. More precisely, $\ImagePhases$ is given by 
\begin{align}\label{Eq:ImagePhases}
  \ImagePhases(x) = \begin{cases}
       i,  \qquad &\text{if } x \text{ belongs to phase }i\in\{1,2,\dots,\NumberPhases\},\\
       0, \qquad &\text{otherwise},
   \end{cases} 
\end{align}
for each $x\in\WindowIntactParticle$. In Figures~\ref{Fig:Visualization particles}b and ~\ref{Fig:Visualization particles}d both composite particles are visualized, where voxels assigned to different phases are visualized by different colors. For both composite particles considered in this paper, we observed three different clusters in the CT image data, where we assume that these clusters correspond to three different mineral phases, i.e., $\NumberPhases=3$. For each $i\in\{1,2,3\}$, the $i$-th phase $P_i$ is then given by $ P_i = \{x\in\WindowIntactParticle:\ImagePhases(x) = i\}$.

Using the phase-based segmentation stated above, a particle can be characterized by aggregated descriptors such as the volume fractions of its mineral phases, which are crucial for quantifying the transgranular and intergranular fracture mechanisms. For each $i\in\{1,2,3\}$, the volume fraction $\VolumePhase_i\in[0,1]$ of the $i$-th phase is given by 
\begin{align}
    \VolumePhase_i = \frac{\lvert P_i\rvert}{\lvert\{x\in\WindowIntactParticle:\ImagePhases(x)>0\}\rvert}.
\end{align}
In our case, both composite particles consist mainly of a single phase, called $P_1$ in the following, which has the largest volume fraction, see Table~\ref{Tab:Phase-specific fracture share}. The slag particle additionally consists of a second phase $P_2$ with around $20\%$ of the particle's volume, whereas the phases $P_2$ and $P_3$ of the ore particle only share very small volume fractions. In the slag particle, phase $P_3$ corresponds to air bubbles, whereas in the ore particle no air bubbles have been detected. Although we denote the three phases of the ore and slag particles by the same symbol (i.e., $P_1, P_2$ and $P_3)$, this does not mean that these phases correspond in both particles to the same mineral.

\begin{table}[ht!]
\caption{Volume fractions of mineral phases in the ore and slag particles, observed in image measurements before fracture.
\label{Tab:Phase-specific fracture share}}  
\newcolumntype{D}{>{\centering}p{0.4\textwidth}}
		\begin{tabularx}{\textwidth}{DCCC}
			\toprule
			&phase $P_1$ (blue) &phase $P_2$ (orange) &phase $P_3$ (green)\\
   \midrule
			$\VolumePhase_i$ (ore)&0.96&0.04&0.001\\
            $\VolumePhase_i$ (slag)&0.78 & 0.22 & 0.006\\
			\bottomrule
		\end{tabularx}
\end{table}

%%%%%%%%%%%%%%%%%%%%%%%%%%%%%%%%%%%%%%%%%%%%%%%%%%%%%%%%%%%%%%%%%%%%%%%%%%%%%%%%%%%%

\subsection{Fragment-based segmentation}\label{Sec:Fragment-based segmentation}
Grayscale images of broken particles do not directly provide a segmentation into individual fragments, because the fragments are partially interconnected, making it impossible to identify single fragments as connected components. Therefore, we employ a combination of the watershed algorithm~\cite{Beucher1993, Beare2006, Roerdink2000, Soille2013} and a 3D morphological reconstruction approach~\cite{Zheng2016, Ridder2023} to segment grayscale images of broken particles into individual fragments while simultaneously preventing oversegmentation. In comparison to techniques which are commonly used in the literature to prevent oversegmentation, like extended regional minima~\cite{Spettl2015} or post-processing steps involving dilation~\cite{Kuchler2018}, the advantage of our approach is that it only requires manual tuning of a single parameter. 
%It turns out that the specific value of $\alpha$ equal $0.1$ is a reasonable choice, and this decision aligns with a visual assessment of the number of fragments observed in the image data.

The output of our segmentation procedure is an image $\watershed \colon \WindowBrokenParticle \to \{0,\dots,\NumberFragments\}$ where $\NumberFragments\geq 1$ denotes the number of fragments and $\watershed$ assigns each voxel to the $i$-th fragment or the background. More precisely, it holds that
\begin{align}
   \watershed(x) = \begin{cases}
       i,  \qquad &\text{if } x \text{ belongs to the $i$-th fragment for some }i\in\{1,2,\dots,\NumberFragments\},\\
       0, \qquad &\text{otherwise},
   \end{cases} 
\end{align}
for each $x\in\WindowBrokenParticle$. Then, we get $\NumberFragments$ fragments $\Fragment_1,\dots,\Fragment_{\NumberFragments}\subset\WindowBrokenParticle$, where $\Fragment_i= \{x\in \WindowBrokenParticle \colon \watershed(x)=i \}$ for each $i=1,\dots, \NumberFragments$. For the purpose of reassembling a fragment $\Fragment \in \SetFragments=\{\Fragment_1,\dots,\Fragment_{\NumberFragments} \}$ in Section~\ref{Sec:Virtual reassembling algorithm}, we need the grayscale value $\ImageBrokenParticle(x)$ for each voxel $x \in \Fragment$. Therefore, we define the fragment image $\ImageFragment \colon \WindowFragment \to \{0,1,\dots,65565\}$ by cropping, i.e., 
\begin{align}
   \ImageFragment(x) = \begin{cases}
       \ImageBrokenParticle(x),  \qquad &\mbox{if $x \in \Fragment$,}\\
       0, \qquad &\mbox{otherwise},
   \end{cases} 
\end{align}
for each $x\in\WindowFragment$, where $\WindowFragment \subset \Z^3$ denotes the smallest cuboidal sampling window which contains the fragment $\Fragment$. Thus, the mapping $\ImageFragment$, assigning each voxel of $x\in\Fragment$ the corresponding grayscale value of $\ImageBrokenParticle$, can be considered to be a cutout of $\ImageBrokenParticle$ representing the fragment $\Fragment$. Furthermore, we consider the mask $\MaskFragment:\WindowFragment\rightarrow\{0,1\}$ of the fragment $\Fragment$, where $\MaskFragment(x)=1$ for each $x\in\Fragment$, and $\MaskFragment(x)=0$ for $x\in\WindowFragment\setminus\Fragment$.

%%%%%%%%%%%%%%%%%%%%%%%%%%%%%%%%%%%%%%%%%%%%%%%%%%%%%%%%%%%%%%%%%%%%%%%%%%%%%%%%%%%%
%%%%%%%%%%%%%%%%%%%%%%%%%%%%%%%%%%%%%%%%%%%%%%%%%%%%%%%%%%%%%%%%%%%%%%%%%%%%%%%%%%%%

\section{Virtual reassembling algorithm}\label{Sec:Virtual reassembling algorithm}

This section outlines the virtual reassembling algorithm which is used to virtually arrange a family $\SetFragments=\{\Fragment_1,\dots,\Fragment_{\NumberFragments}\}$ of fragment images to a new image, reassembling the reference image $\ImageIntact$ of a non-broken particle. In the reassembling process, the fragment image $\ImageFragment$ is iteratively registered by aligning $\ImageFragment$ with the reference image $\ImageIntact$ for each $\Fragment\in\SetFragments$. This alignment involves finding a transformation that minimizes the discrepancy between the transformed fragment image and the reference image of the non-broken particle. More precisely, we determine the translation and rotation of fragment images $\ImagesSetFragments=\{\ImageFragment\}_{\Fragment\in\SetFragments}$ with masks $\MaskSetFragments=\{\MaskFragment\}_{\Fragment\in\SetFragments}$, taking into account that during fracture, a fragment $\Fragment\in\SetFragments$ may shift and rotate away from its original position in the non-broken particle, where the transformation of a fragment image $\ImageFragment$, which aligns it within $I_\mathrm{intact}$, is determined by applying a so-called mask-based image registration approach. In the present paper, we use an intensity-based image registration approach~\cite{Barnea1972} extended with PSO~\cite{Kennedy1995, Kennedy2006}, which has been successfully utilized in biomedical image registration~\cite{Wachowiak2004} for the registration of single pairs of 3D image data. We extend this approach by not only registering two images, but by proposing a reassembly algorithm, which iteratively registers the images of a family of (possibly multiple) fragment images with a reference image that is updated during reassembly. Furthermore, we consider masking in the image registration approach by applying FFT as described by Padfield~\cite{Padfield2012}.

The rest of this section is divided into two parts. In Section~\ref{Sec:Mask-based image registration}. to make our paper more self-contained, the mask-based image registration approach is explained for a general setup. Then, in Section~\ref{Sec:Virtual reassembling of fragments} we describe how the reassembling algorithm iteratively utilizes the mask-based image registration approach in order to reassemble a set of fragments.

%%%%%%%%%%%%%%%%%%%%%%%%%%%%%%%%%%%%%%%%%%%%%%%%%%%%%%%%%%%%%%%%%%%%%%%%%%%%%%%%%%%%

\subsection{Mask-based image registration}\label{Sec:Mask-based image registration}

In order to determine the translation and rotation of an image with respect to a reference image, image registration algorithms can be utilized~\cite{Wachowiak2004}, where a moving image $\ImageMoving:\WindowImageMoving\rightarrow\{0,\dots,65535\}$ (corresponding to a fragment image $\ImageFragment$) and a fixed image $\ImageFixed:\WindowImageFixed\rightarrow\{0,\dots,65535\}$ (corresponding to the image $I_\mathrm{intact}$ of the non-broken particle) are registered. Here, the fixed image $\ImageFixed$ is observed within a certain sampling window $\WindowImageFixed\subset\Z^3$, and the moving image $\ImageMoving$ within $\WindowImageMoving\subset\Z^3$. 

An important step in implementing image registration algorithms is to determine a suitable function $T:\R^3\rightarrow \R^3$, which maps a point $x^{\mathrm{m}}\in\R^3$ to a point $x^{\mathrm{f}}\in\R^3$, i.e., $T(x^{\mathrm{m}}) = x^{\mathrm{f}}$. Such functions can be used to define transformed images, where, in the context of image registration, the transformed moving image should resemble the fixed image. As we consider translation and rotation operations in the registration process, the function $T$ is a superposition of a  function $\ShiftFunction:\R^3\rightarrow\R^3$ which shifts a point $x\in\R^3$ by a translation vector $t\in\R^3$ to the new position $x+t\in\R^3$, and a function $\RotationFunction:\R^3\rightarrow\R^3$ which rotates a point $x\in\R^3$ by a vector of Euler angles $\theta=(\theta_1,\theta_2,\theta_3)^{}\in[0,2 \pi) \times [0,\pi) \times [0, 2 \pi)$, see~\cite{Burger2016} for details. Note that singularities which might occur by using Euler angles are not addressed here, instead we refer e.g. to~\cite{Diebel2006}. The function $T=\Transformationfunction:\R^3\to\R^3$ is then given by
\begin{align}\label{Eq:TransformationFunction}
    \Transformationfunction(x) = \RotationFunction(\ShiftFunction(x)),
\end{align}
for each $x\in\R^3$. Thus, the transformed moving image $\transformedImage \colon \WindowImageFixed \rightarrow \{0,\dots,65535\}$ is given by
\begin{align}\label{Eq:TransformatedImage}
\transformedImage(x)=\ImageMoving(\Transformationfunction^{-1}(x)),
\end{align}
for each $x\in \WindowImageFixed$. If, for some $x\in \WindowImageFixed$, the value of $\Transformationfunction^{-1}(x)$ on the right-hand side of Eq.~(\ref{Eq:TransformatedImage}) does not belong to the convex hull of $\WindowImageMoving$, we put $\transformedImage(x)=0$. Otherwise, since the mapping $\ImageMoving$ is defined on the discrete set $\WindowImageMoving\subset\Z^3$, whereas $\Transformationfunction^{-1}$ can take arbitrary values in the (continuous) Euclidean space $\R^3$, we employ spline interpolation of order zero such that $\Transformationfunction^{-1}(x)$ is replaced by the closest voxel of $\WindowImageMoving$~\cite{Burger2016}.

% of image intensity values for computing the transformed image, i.e., if $\Transformationfunction^{-1}(x)\notin\WindowImageMoving$ for $x\in\WindowImageFixed$, then we compute the value $\ImageMoving(\Transformationfunction^{-1}(x))$ by spline interpolation of order zero~\cite{Burger2016}. 

% The goal of the registration approach is to determine the shift $t\in\R^3$ of the shift function $\ShiftFunction$ and the rotation parameter $\theta\in[0,2 \pi) \times [0,\pi) \times [0, 2 \pi)$ of the rotation function $\RotationFunction$ for which $\transformedImage$ looks similar to $\ImageFixed$. Thus, 

The objective of image registration is to maximize the similarity between the transformed moving image $\transformedImage$ and the fixed image $\ImageFixed$, where similarity is quantified by a so-called similarity measure which we will further specify in Section~\ref{Sec:Similarity measure} below. Therefore, the image registration problem can be reformulated as an optimization problem with some objective function $f:\R^3\times[0,2 \pi) \times [0,\pi) \times [0, 2 \pi)\rightarrow[0,1]$, which maps parameter pairs of shift and rotation, $t\in \R^3$ and $\theta \in [0,2 \pi) \times [0,\pi) \times [0, 2 \pi)$, onto the similarity $f(t,\theta)$ between $\transformedImage$ and $\ImageFixed$. The goal is to maximize $f(t,\theta)$. In other words, we seek an optimal shift parameter $t_{\mathrm{opt}}\in\R^3$ and an optimal rotation parameter $\theta_{\mathrm{opt}}\in[0,2 \pi) \times [0,\pi) \times [0, 2 \pi)$ such that
    \begin{align}\label{Eq:GlobalOptimizationProblem}
        (t_{\mathrm{opt}},\theta_{\mathrm{opt}}) = \argmax_{t\in\R^3,\theta\in[0,2 \pi) \times [0,\pi) \times [0, 2 \pi)} f(t,\theta).
\end{align}
It is important to note that maximizing $f(t,\theta)$ with respect to both parameters $t$ and $\theta$ simultaneously can be challenging in practice due to the complexity of the optimization problem stated in Eq.~(\ref{Eq:GlobalOptimizationProblem}), involving six degrees of freedom (three shift coordinates and three rotations around the three axes). Therefore, we adopt an iterative approach in an inner-outer circle step fashion. In the inner circle step, for a given $\theta$, we determine an optimal value $t_\theta$ for the shift parameter $t$—the one that maximizes $f(\cdot,\theta)$. In the outer circle step, we then determine an optimal value for the rotation parameter $\theta$ which maximizes $f( t_\theta,\theta)$. %This iterative process helps to overcome the complexity of the optimization problem.

To determine an optimal shift $t_\theta\in\R^3$ for a given rotation parameter $\theta$, we employ a mask-based image registration procedure which is implemented via $\FFT$, as outlined in~\cite{Padfield2012}. Note that in the context of reassembling fragments, it is essential to apply masking to both (moving and fixed) images. More precisely, masking is crucial because a fragment $\Fragment\in\SetFragments$ within the 3D image $\ImageBrokenParticle$ of the broken particle is often surrounded by background or even adjacent partially observable fragments. Thus, image registration without using the mask of a fragment could wrongly be influenced by such regions outside of $\Fragment$. Additionally, registering the fragment within the image $\ImageIntact$ of the non-broken particle requires masking of the latter. This ensures that the moving image is registered within the non-broken particle in the fixed image. In particular, by $\MaskMoving:\WindowImageMoving\rightarrow\{0,1\}$ and $\MaskFixed:\WindowImageFixed\rightarrow\{0,1\}$ we denote the masks of the moving and fixed image, respectively. Then, the mask $\MaskMoving_{t,\theta}:\WindowImageFixed\rightarrow\{0,1\}$ of the transformed moving image $\transformedImage$ introduced in Eq.~(\ref{Eq:TransformatedImage}) is given by 
\begin{align}
   \MaskMoving_{t,\theta}(x) = \MaskMoving(\Transformationfunction^{-1}(x)),   
\end{align}
for each $x\in \WindowImageFixed.$ Furthermore, we consider the set
\begin{align}\label{Eq:Restiction}
     D_{t,\theta} = \{x \in \WindowImageFixed : \MaskFixed(x)=\MaskMoving_{t,\theta}(x)=1\},    
\end{align}
% \begin{align}
%     D^\mathrm{f}=\{y\in\WindowImageFixed:\MaskFixed(y)=1\},\,  D^\mathrm{m}=\{y\in\WindowImageMoving:\MaskMoving(y)=1\} \text{ and } D_{t,\theta} = \{y \in \WindowImageFixed : \MaskFixed(x)=\MaskMoving_{t,\theta}(x)=1\},    
% \end{align}
for any shift $t\in\R^3$ and orientation angle vector $\theta \in [0,2 \pi) \times [0,\pi) \times [0, 2 \pi)$, where $D_{t,\theta}$ contains those foreground voxels of the fixed image $\ImageFixed$ which are foreground voxels of the transformed moving image $\transformedImage$. 

%In the subsequent sections, first the similarity measure is introduced in Section~\ref{Sec:Similarity measure} to assess the quality of the image registration result. Secondly, the optimization of the similarity measure is described in Section~\ref{Sec:Optimization of the similarity measure}.

%%%%%%%%%%%%%%%%%%%%%%%%%%%%%%%%%%%%%%%%%%%%%%%%%%%%%%%%%%%%%%%%%%%%%%%%%%%%%%%%%%%%

\subsubsection{Similarity measure}\label{Sec:Similarity measure}
To assess the similarity between a transformed moving image $\transformedImage$ and the fixed image $\ImageFixed$, restricted to the set $D_{t,\theta}$ given in Eq.~(\ref{Eq:Restiction}), a similarity measure is required to indicate how closely the grayscale values of $\transformedImage$ match those of $\ImageFixed$. In this context, the masked normalized cross-correlation coefficient (MNCC) is employed. Note that the normalized cross-correlation coefficient (NCC) is a widely used similarity measure in image registration for assessing registration outcomes~\cite{Burger2016}. However, in the present paper, we consider a modified version of the NCC, which is adapted to masked images~\cite{Padfield2012}. More precisely, we measure the similarity of the moving image $\ImageMoving$, transformed by $\Transformationfunction$, and the fixed image $\ImageFixed$ by means of the quantity $\MNCC_{t,\theta}(\ImageFixed,\ImageMoving) \in [-1,1]$, which is given by
\begin{align}\label{Eq:Definition MNCC}
    \MNCC_{t,\theta}(\ImageFixed,\ImageMoving) = \frac{\sum_{x\in D_{t,\theta}}(\ImageFixed(x)-\MeanImageFixed)(\transformedImage(x)-\MeanImageMoving)}{\sqrt{\sum_{x\in D_{t,\theta}}(\ImageFixed(x)-\MeanImageFixed)^2}\sqrt{\sum_{x\in D_{t,\theta}}(\transformedImage(x)-\MeanImageMoving)^2}},
\end{align}
for all $t\in\R^3$ and $\theta\in[0,2 \pi) \times [0,\pi) \times [0, 2 \pi)$, where  values of $\MNCC_{t,\theta}(\ImageFixed,\ImageMoving)$ close to $1$ indicate a large similarity of $\ImageMoving$ and $\ImageFixed$. The mean values $\MeanImageFixed$ and $\MeanImageMoving$ are given by
\begin{align}\label{Eq:Mean values}
    \MeanImageFixed = \frac{1}{\lvert D_{t,\theta}\rvert}\sum_{x\in D_{t,\theta}}\ImageFixed(x)\qquad \text{and}\qquad \MeanImageMoving = \frac{1}{\lvert D_{t,\theta}\rvert}\sum_{x\in D_{t,\theta}} \transformedImage(x).
\end{align}
Note that in the case of $D_{t,\theta} = \varnothing$ we put $\MNCC_{t,\theta}(\ImageFixed,\ImageMoving)=-1$, i.e., to the smallest similarity. The evaluation of the sums in Eq.~(\ref{Eq:Definition MNCC}) can be computationally expensive, especially in the optimization process, since they must be repeatedly computed for all considered candidates for $t$ and $\theta$. To address this challenge, we represent all these sums directly in the Fourier domain, following the method outlined by Padfield~\cite{Padfield2012}. This approach involves treating the sums in Eq.~(\ref{Eq:Definition MNCC}) as convolutions which corresponds to point-wise multiplication in the Fourier domain, where the latter can be efficiently computed using FFT. This enables an efficient computation of $\MNCC_{t,\theta}(\ImageFixed,\ImageMoving)$ for any potential shift $t\in\R^3$ and for any given rotation $\theta\in[0,2 \pi) \times [0,\pi) \times [0, 2 \pi)$.

By using the quantity
 $\MNCC_{t,\theta}(\ImageFixed,\ImageMoving)$ considered in Eq.~(\ref{Eq:Definition MNCC})
as similarity measure, the objective function $f$ to be optimized in Eq.~(\ref{Eq:GlobalOptimizationProblem}) is given by
\begin{align}
    f(t,\theta) = \MNCC_{t,\theta}(\ImageFixed,\ImageMoving) 
\end{align}
for any $t\in\R^3$ and $\theta\in[0,2 \pi) \times [0,\pi) \times [0, 2 \pi)$.

%%%%%%%%%%%%%%%%%%%%%%%%%%%%%%%%%%%%%%%%%%%%%%%%%%%%%%%%%%%%%%%%%%%%%%%%%%%%%%%%%%%%

\subsubsection{Optimization of shift and rotation}\label{Sec:Optimization of the similarity measure}
As mentioned above, for any fixed rotation $\theta\in[0,2 \pi) \times [0,\pi) \times [0, 2 \pi)$, the optimal shift $t_{\mathrm{opt},\theta} \in \R^3$ such that 
\begin{align}
    t_{\mathrm{opt},\theta} = \argmax_{t \in \R^3} f(t,\theta)
\end{align}
can be computed efficiently using the $\FFT$-based method outlined in~\cite{Padfield2012}.
Therefore, the optimization problem considered in Eq.~(\ref{Eq:GlobalOptimizationProblem}) can be written as
\begin{align}\label{Eq:NewOptimizationProblem}
    \theta_\mathrm{opt}=\argmax_{\theta \in [0,2 \pi) \times [0,\pi) \times [0, 2 \pi)} f( t_{\mathrm{opt},\theta},\theta),
    \quad \text{and} \quad t_\mathrm{opt} = t_{\mathrm{opt}, \theta_\mathrm{opt}}.
\end{align}
Thus, in Eq.~(\ref{Eq:NewOptimizationProblem}) we only have to consider an optimization problem with three variables instead of one with six, as it is the case in Eq.~(\ref{Eq:GlobalOptimizationProblem}).

In order to numerically solve the optimization problem formulated in Eq.~(\ref{Eq:NewOptimizationProblem}), we deploy PSO~\cite{Kennedy1995, Kennedy2006}. This method, which turned out to be successful in biomedical image registration~\cite{Wachowiak2004}, will be adapted for the present application of reassembling fragments. This means in particular that the optimization algorithm operates under minimal assumptions on the objective function being optimized and can search large spaces of candidates. Unlike classical optimization methods, such as gradient descent and quasi-Newton methods, PSO does not rely on the gradient of the objective function. Induitively speaking, the PSO algorithm considers numerous candidates (called particles in this context) for solving the optimization problem which it updates iteratively. In our application, i.e., for solving the optimization problem formulated in Eq.~(\ref{Eq:NewOptimizationProblem}), we put the total number of particles equal to $3$.

As output of this procedure we obtain an optimal pair of rotation $\theta_\mathrm{opt}$ and shift $t_\mathrm{opt}$, which can then be used to obtain the optimally transformed moving image $\transformedImageOptimal:\WindowImageFixed\rightarrow\{0,1,\dots,65565\}$ given by
\begin{align}\label{Eq:OptimalTransformedMovingImage}
\transformedImageOptimal(x)=\ImageMoving(\TransformationfunctionOptimal^{-1}(x)),
\end{align}
for each $x\in \WindowImageFixed$. Furthermore, we obtain the optimally transformed moving mask $\transformedMaskOptimal:\WindowImageFixed\rightarrow\{0,1\}$ given by
\begin{align}\label{Eq:OptimalTransformedMovingMask}
\transformedMaskOptimal(x)=\MaskMoving(\TransformationfunctionOptimal^{-1}(x)),
\end{align}
for each $x\in \WindowImageFixed$.

%%%%%%%%%%%%%%%%%%%%%%%%%%%%%%%%%%%%%%%%%%%%%%%%%%%%%%%%%%%%%%%%%%%%%%%%%%%%%%%%%%%%

\subsection{Virtual reassembling of fragments}\label{Sec:Virtual reassembling of fragments}

This section describes the virtual reassembling algorithm that is applied to a family of fragments $\SetFragments=\{\Fragment_1,\dots,\Fragment_{\NumberFragments}\}$ represented by the images $\ImagesSetFragments=\{I_{\Fragment_{i}}\}_{i=1}^{\NumberFragments}$ and masks $\MaskSetFragments=\{M_{\Fragment_{i}}\}_{i=1}^{\NumberFragments}$. Simply speaking, the algorithm works iteratively by reassembling one fragment in each step before continuing with the next fragment. In this context, the reassembling of a fragment $\Fragment\in\SetFragments$ can be achieved by considering $\ImageFragment$ as the moving image with mask $\MaskFragment$ and $I_\mathrm{intact}$ as the fixed image with mask $\MaskImageIntact$, followed by applying the image registration method described in Section~\ref{Sec:Mask-based image registration}. In each step of the iterative reassembling algorithm, the reassembling results of the previous steps are taken into account. Therefore, in each step the fixed image $\ImageFixed$ used for image registration is updated by removing the fragment registered in the current step from the fixed image. In this manner, further fragments tend to be moved/rotated to regions of the fixed image in which no fragments have been placed in previous steps. By incorporating this modification into the algorithm, the outcome of the reassembling algorithm depends on the quality of the reassembled fragment at each step. Consequently, we introduce a similarity measure in Section~\ref{Sec:Evaluation step}, which is then used in each step of the algorithm to assess the quality of a reassembled fragment. Furthermore, it has turned out that the order of fragments being reassembled influences the reassembling result. Thus, we provide a strategy on how to reassemble fragments. Overall, the reassembling algorithm is designed as sketched below.

\begin{algorithm}[ht!]
    \caption{\label{algo:VirtualReassemblingAlgorithm}Virtual reassembling algorithm}
    \begin{algorithmic}
        \Function{VirtualReassembleFragments}{Set of fragments $\SetFragments$, images of fragments $\ImagesSetFragments$, masks of fragments $\MaskSetFragments$, image of non-broken particle $\ImageIntact$, mask of non-broken particle $\MaskImageIntact$}
            \State $\{\SetFragments_i\}_{i=1}^{n}\gets$ FragmentSetPartition($\SetFragments$) as described in Section~\ref{Sec:Virtual reassembling order}
            \State $\ImageFixed\gets\ImageIntact$ and $\MaskFixed\gets\MaskImageIntact$
            \State Set $\ImageReassemblingResultLabeled:\WindowIntactParticle\rightarrow \{0\}$
            \ForAll{$i=1,2,\dots,n$}
                \State $\MeanSSM_{\mathrm{best}} \gets -1$
                \While{$\SetFragments_i \neq\emptyset$}
                \ForAll{fragments $\Fragment\in \SetFragments_i$}
                    \State $\ImageMoving \gets \ImageFragment$ and $\MaskMoving \gets \MaskFragment$
                    \State $\transformedImageOptimal$ and $\transformedMaskOptimal \gets$ ImageRegistration($\ImageFixed,\MaskFixed,\ImageMoving,\MaskMoving$) as described in Section~\ref{Sec:Mask-based image registration}
                    \State $\ImageFixedAdjusted \gets$ ModifyFixedImage($\ImageFixed,\MaskFixed,\transformedImageOptimal,\transformedMaskOptimal$) by means of Eq.~(\ref{Eq:AdjustedFixedImage for MSSM})
                    \State $\MeanSSM_{\Fragment}\gets$ ComputeMSSM($\ImageFixed, \ImageFixedAdjusted,\transformedMaskOptimal$) by means of Eq.~(\ref{Eq:Mean structural similarity measure})
                    \If{$\MeanSSM_{\Fragment} \geq \MeanSSM_{\mathrm{best}}$}
                        \State $\widetilde{\Fragment}\gets \Fragment$, $M_{\widetilde{\Fragment}}\gets \transformedMaskOptimal$ and $\MeanSSM_{\mathrm{best}}\gets \MeanSSM_{\Fragment}$
                    \EndIf
                \EndFor
                \State $\SetFragments_i\gets \SetFragments_i\setminus \{\widetilde{\Fragment}\}$
                \State $\ImageFixed$ and $\MaskFixed \gets $ UpdateFixedImage($\ImageFixed,\MaskFixed,M_{\widetilde{\Fragment}}$) by means of Eq.~(\ref{Eq:AdjustedFixedImage}) and~(\ref{Eq:AdjustedFixedMask})
                \State $\ImageReassemblingResult \gets$ UpdateReassemblingResult($\ImageReassemblingResult,M_{\widetilde{\Fragment}}$) by means of Eq.~(\ref{Eq:Update Reassembling result labeled})
                \EndWhile
            \EndFor
        %\State $\ImageReassemblingResultLabeled \gets$ DetermineReassemblingResultLabeled($\ImageReassemblingResultLabeled$) by means of Eq.~(\ref{Eq:Update Reassembling result labeled})\\
        \Return $\ImageReassemblingResultLabeled$
        \EndFunction
    \end{algorithmic}
\end{algorithm}

 We now describe the similarity measure for evaluating how well we reassembled a fragment, see Section~\ref{Sec:Evaluation step}. Afterwards, we explain the process of modifying the fixed image, which is used for registration after each step of the reassembling algorithm, see Section~\ref{Sec:Updating step}. Then, Section~\ref{Sec:Virtual reassembling order} outlines the virtual reassembling order, specifying the particular set of fragments from which the fragment to be reassembled is chosen. Finally, in Section~\ref{Sec:Final output of the reassembling algorithm}, we describe the kind of results obtained by the reassembling algorithm.

%%%%%%%%%%%%%%%%%%%%%%%%%%%%%%%%%%%%%%%%%%%%%%%%%%%%%%%%%%%%%%%%%%%%

\subsubsection{Evaluation of reassembling quality}\label{Sec:Evaluation step}

Typically, there are multiple fragments which have to be reassembled. Thus, we need a strategy for choosing the fragment which, after reassembly, leads to the best result. It turned out that the MNCC given in Eq.~(\ref{Eq:Definition MNCC}) is an appropriate similarity measure in mask-based image registration of single fragments. However, for evaluating the reassembling results of multiple fragments we use the mean structural similarity measure ($\MeanSSM$) as described in~\cite{Wang2004}. By utilizing $\MeanSSM$, we can measure the structural similarity of two images by comparing the local patterns of their voxel values which have been normalized for luminance and contrast. This similarity measure is advantageous for evaluating the reassemble results of different fragments and it is expected to have a larger value if a fragment is better reassembled than another fragment. In the following, we state the definition of $\MeanSSM$ and show how this measure is applied to evaluate the goodness of fit of a reassembled fragment.

Recall that for each fragment $\Fragment\in\SetFragments$ with image $\ImageFragment$ and mask $\MaskFragment$, we registered the moving image $\ImageMoving=\ImageFragment$ within the fixed image $\ImageFixed=\ImageIntact$, using the mask-based registration procedure described in Section~\ref{Sec:Mask-based image registration}, where the registration leads an (optimally) transformed moving image $\transformedImageOptimal$ with mask $\transformedMaskOptimal$, see Eqs.~(\ref{Eq:OptimalTransformedMovingImage}) and~(\ref{Eq:OptimalTransformedMovingMask}).
% for a fragment $\Fragment\in\SetFragments$, the mask-based registration, as described in Section~\ref{Sec:Mask-based image registration}, results in an optimally transformed moving image $\transformedImageOptimal$ with optimally transformed mask $\transformedMaskOptimal$, where $\transformedImageOptimal$ corresponds to the image $\ImageFragment$ of $F$ and the fixed image $\ImageFixed$ corresponds to the image $\ImageIntact$ of the non-broken particle. 
Now, in order to evaluate the goodness of fit of the masked-based registration result for $\Fragment$, we utilize the image $\ImageFixedAdjusted:\WindowImageFixed\rightarrow\{0,1,\dots,65565\}$, which is given by 
\begin{align}\label{Eq:AdjustedFixedImage for MSSM}
       \ImageFixedAdjusted(x) = \begin{cases}
           \transformedImageOptimal(x),\qquad & \text{if } \transformedMaskOptimal(x)=1,\\
            \ImageFixed(x),\qquad & \text{otherwise},
        \end{cases}
    \end{align}
for each $x\in\WindowImageFixed$. Note that $\ImageFixedAdjusted$ coincides with the mask-based registration result $\transformedImageOptimal$ within its mask $\transformedMaskOptimal$, where $\ImageFixedAdjusted$ is compared to $\ImageFixed$ using $\MeanSSM$. More precisely, we evaluate the structural similarity of the images $\ImageFixed$ and $\ImageFixedAdjusted$, restricted to the set
$\BoundingBoxtransformedMaskOptimal = \{x \in \WindowImageFixed : \transformedMaskOptimal(x)=1\}$,
 by means of
  \begin{align}\label{Eq:Mean structural similarity measure}
        \MeanSSM(\ImageFixed, \ImageFixedAdjusted, \transformedMaskOptimal) = \frac{1}{\lvert \BoundingBoxtransformedMaskOptimal\rvert}\sum_{x\in \BoundingBoxtransformedMaskOptimal}\SSM_{\ImageFixed,\ImageFixedAdjusted}(x),
\end{align} 
where the formal definition of $\SSM$ is given in the Appendix. Here, we just remark that $\SSM_{\ImageFixed,\ImageFixedAdjusted}(x)$ is a structural similarity measure of $\ImageFixed$ and $\ImageFixedAdjusted$ evaluated at $x\in\BoundingBoxtransformedMaskOptimal$. Moreover, the definition of $\SSM$ immediately implies that 
$\MeanSSM(\ImageFixed, \ImageFixedAdjusted, \transformedMaskOptimal)\in[0,1]$, where
values of $\MeanSSM(\ImageFixed, \ImageFixedAdjusted, \transformedMaskOptimal)$ close to $1$ indicate a high degree of structural similarity. In particular, in the context of reassembling, a value of $\MeanSSM(\ImageFixed, \ImageFixedAdjusted, \transformedMaskOptimal)$ close to $1$ indicates that the fragment $\Fragment$ is well reassembled. 
%Here, $\SSM$ denotes the. 
%More details regarding $\SSM$ are given in the Appendix
%~\ref{appendix:A}. 

%%%%%%%%%%%%%%%%%%%%%%%%%%%%%%%%%%%%%%%%%%%%%%%%%%%%%%%%%%%%%%%%%%%%%%%%%%%%%%%%%%%%

\subsubsection{Updating the fixed image}\label{Sec:Updating step}

In this section, we explain how the fixed image $\ImageFixed$, used in the mask-based registration procedure, is updated after identifying the best reassembled fragment $\Fragment\in\SetFragments$, i.e., that fragment $\Fragment\in\SetFragments$ which leads to the largest value of $\MeanSSM$ in comparison to all other fragments. Note that without updating $\ImageFixed$, 
we would not take the reassembling result of $\Fragment$ into account in subsequent steps of the reassembling algorithm, which could lead to overlapping of reassembled fragments.

%Therefore, $\ImageFixed$ is updated after each step of the virtual reassembling algorithm, where the updating works as follows: 

More precisely, we update $\ImageFixed$ as follows: After identifying the best reassembled fragment $\Fragment\in\SetFragments$, with (optimally) transformed image $\widetilde{I}_{\Fragment}$ and mask $\widetilde{M}_{\Fragment}$ determined by means of Eqs.~(\ref{Eq:OptimalTransformedMovingImage}) and~(\ref{Eq:OptimalTransformedMovingMask}), respectively, we replace $\ImageFixed$ by the image $\ImageFixedAdjusted:\WindowImageFixed\rightarrow\{0,1,\dots,65565\}$, which is given by
    \begin{align}\label{Eq:AdjustedFixedImage}
        \ImageFixedAdjusted(x) = \begin{cases}
            0,\qquad & \text{if } \widetilde{M}_{\Fragment}(x)=1,\\
            \ImageFixed(x),\qquad & \text{otherwise},
        \end{cases}
    \end{align}
for each $x\in \WindowImageFixed$. Analogously, we replace $\MaskFixed$ by the image $\MaskFixedAdjusted:\WindowImageFixed\rightarrow\{0,1\}$, where 
\begin{align}\label{Eq:AdjustedFixedMask}
        \MaskFixedAdjusted(x) = \begin{cases}
           0,\qquad & \text{if } \widetilde{M}_{\Fragment}(x)=1,\\
           \MaskFixed(x),\qquad & \text{otherwise},
        \end{cases}
    \end{align}
for each $x\in\WindowImageFixed$. By incorporating $\ImageFixedAdjusted$ as the new fixed image during registration of a fragment image in the next step of the algorithm, we take into consideration the outcomes of previous steps. This approach effectively constrains the range of potential translations and rotations for reassembling fragments in the following steps of the reassembling algorithm.

%%%%%%%%%%%%%%%%%%%%%%%%%%%%%%%%%%%%%%%%%%%%%%%%%%%%%%%%%%%%%%%%%%%%%%%%%%%%%%%%%%%%

\subsubsection{Reassembling order}\label{Sec:Virtual reassembling order}

We observed that choosing a random order in which we register fragment images does not lead to satisfactory reassembling results. A reason for this is that fragments can vary in terms of their size (fragment volume), shape and original position within the non-broken particle. The reassembling process for fragments entirely contained within the non-broken particle relies solely on the texture exhibited in the interior of the non-broken particle. In contrast, the reassembling process for fragments located at the boundary of the non-broken particle is influenced by both texture and the shape of the boundary, i.e., further information is available which in turn makes registration easier. Moreover, an additional challenge arises when reassembling small fragments. These smaller fragments often exhibit a more homogeneous texture, leading to a wide range of appropriate translations and rotations within the registration step of the algorithm which results in a good reassembling quality.

Therefore, we introduce a heuristics with which we prioritize the registration of ``promising" fragments. Then, due to the subsequent updates of the fixed image, the set of potential translations and rotations decreases in size, see Section~\ref{Sec:Updating step}. In other words, this leads to less ambiguity when registering more difficult, low priority cases like images of small fragments. More precisely, we partition the set $\SetFragments$ of all fragments into $n\ge 1$ pairwise disjoint subsets $\SetFragments_1,\dots,\SetFragments_{n}\subset\SetFragments$ with $\SetFragments= \bigcup_{i=1}^{n}\SetFragments_i$, where each subset $\SetFragments_i$ contains $n^{(i)}\ge 1$ fragments for $i\in\{1,2,\dots,n\}$. 
The specification of this partitioning influences the quality of reassembling results, where partitioning of the set  $\SetFragments$ of fragments based on their sizes turned out to be a reasonable choice, as larger fragments are more likely to share their boundary with the boundary of the non-broken particle. Therefore, it is reasonable to reassemble larger fragments first, i.e., the subsets $\SetFragments_1,\dots,\SetFragments_{n}$ of $\SetFragments$ are defined such that $\SetFragments_1$ contains the largest $n^{(1)}$ fragments in $\SetFragments$, the subset $\SetFragments_2$ contains the largest $n^{(2)}$ fragments in $\SetFragments\setminus\SetFragments_1$, and so on. This procedure is repeated until all fragments of $\SetFragments$ are assigned to some set $\SetFragments_1,\dots,\SetFragments_{n}$.
The number $n$ of subsets and their sizes $n^{(1)},\ldots,n^{(n)}$ depend on the number and sizes of fragments considered in the given virtual reassembling task.  Note that in the case of the experimental data from both the ore and slag particles described in Section~\ref{Sec:Material and sample preparation}, where the number of fragments is small, no partitioning of $\SetFragments$ is required. However, in the simulation study considered in Section~\ref{Sec:Simulation study} we have a larger number of fragments and, thus, we assign three fragments to each set $\SetFragments_1,\dots,\SetFragments_{n-1}$, and the remaining fragments to $\SetFragments_{n}$ . 

In other words, the reassembling algorithm works as follows: First, all fragments of $\SetFragments_1$ are reassembled and the $\MeanSSM$ is computed for each $\Fragment\in\SetFragments_1$
 as described in Section~\ref{Sec:Evaluation step}. Second, the fragment $\widetilde{\Fragment}\in\SetFragments_1$ with the largest $\MeanSSM$ value is identified. The fixed image is then updated by the reassembling result of $\widetilde{\Fragment}$, as described in Section~\ref{Sec:Updating step}, Then, this process is repeated on the set $\SetFragments_1\setminus\{\widetilde{\Fragment}\}$ with the updated fixed image. Once all fragments in $\SetFragments_1$ are reassembled, the algorithm proceeds to the next set $\SetFragments_2$ and continues this process until all fragments are reassembled.

%%%%%%%%%%%%%%%%%%%%%%%%%%%%%%%%%%%%%%%%%%%%%%%%%%%%%%%%%%%%%%%%%%%%%%%%%%%%%%%%%%%%

\subsubsection{Final output of the reassembling algorithm}\label{Sec:Final output of the reassembling algorithm}

The reassembling algorithm creates an image $\ImageReassemblingResultLabeled:\WindowIntactParticle\rightarrow\{0,1,\dots,\NumberFragments\}$, which assigns each voxel of $\WindowIntactParticle$ to a reassembled fragment $\Fragment_i\in\SetFragments$ for some $i\in\{1,2,\dots,\NumberFragments\}$, or to the background, where $\ImageReassemblingResultLabeled$ is constructed iteratively and initialized as $\ImageReassemblingResultLabeled\equiv 0$. After each reassembling step, when we identified the best reassembled fragment $\Fragment_i\in\SetFragments$ for some $i\in\{1,2,\dots,\NumberFragments\}$ with (optimally) transformed image $I_{\Fragment_i}$ and mask $M_{\Fragment_i}$ determined by means of Eqs.~(\ref{Eq:OptimalTransformedMovingImage}) and~(\ref{Eq:OptimalTransformedMovingMask}), respectively, we replace $\ImageReassemblingResultLabeled$ by the image $\ImageReassemblingResultLabeledAdjusted:\WindowIntactParticle\rightarrow\{0,1,\dots,\NumberFragments\}$, which is given by
\begin{align}\label{Eq:Update Reassembling result labeled}
   \ImageReassemblingResultLabeledAdjusted(x) = \begin{cases}
        i,\qquad & \text{if } M_{\Fragment_i}(x)=1,\\
        \ImageReassemblingResultLabeled(x), \qquad &\text{otherwise,}
    \end{cases}
\end{align}
for each $x\in\WindowIntactParticle$. In Section~\ref{Sec:Fracture characterization}, the finally obtained image $\ImageReassemblingResultLabeled$ is used to determine the fracture surfaces.

\section{Simulation study}\label{Sec:Simulation study}

In order to evaluate the reassembling algorithm described in Section~\ref{Sec:Virtual reassembling algorithm}, we applied the algorithm to various sets of virtually generated fragments.  Below we explain the simulation study we conducted for this purpose. In Section~\ref{Sec:Virtual generation of fragments}, we describe the virtual generation of fragments and, subsequently, in Section~\ref{Sec:Evaluation step algorithm for virtual generated fragments} we introduce several evaluation measures to assess the goodness of fit of the reassembling algorithm, considering both the overall reassembling outcome and, in particular, the fit of the fracture surfaces.

%%%%%%%%%%%%%%%%%%%%%%%%%%%%%%%%%%%%%%%%%%%%%%%%%%%%%%%%%%%%%%%%%%%%%%%%%%%%%%%%%%%%

\subsection{Virtual generation of fragments}\label{Sec:Virtual generation of fragments}
In the following, we provide a brief overview of the procedure involved in generating virtual fragments by employing (distance-based) Voronoi tessellations~\cite{Moller1994,Okabe2000}. Such tessellations can be used to partition the sampling window $\WindowIntactParticle$ of the non-broken particle, with image $\ImageIntactParticle$ and mask $\MaskImageIntact$, into a collection of $\NumberTessellationCells\ge 1$ non-overlapping sets $C_1, C_2,\dots,C_{\NumberTessellationCells}\subset \WindowIntactParticle$, so-called cells. This model is fully specified by a point pattern $\mathcal{P}=\{s_i\}^{\NumberTessellationCells}_{i=1}\subset\R^3$, where $s_i\in\R^3$ is the seed point of cell $C_i$ for each $i\in\{1,2,\dots,\NumberTessellationCells\}$. More precisely, the $i$-th cell $C_i$ of the Voronoi tessellation $\Tessellation=\{C_i\}_{i=1}^{\NumberTessellationCells}$ is given by 
\begin{align}\label{Eq:Voronoi}
    C_i = \{x\in \WindowIntactParticle: \lVert x-s_i\rVert^2\leq \lVert x-s_j\rVert^2 \text{ for each } j=1,\dots,\NumberTessellationCells, j\neq i\},
\end{align}
where $\lVert\cdot\rVert$ denotes the Euclidean norm in $\R^3$.

For the purpose of generating virtual fragments, suitable point patterns $\mathcal{P}=\{s_i\}^{\NumberTessellationCells}_{i=1}$ have to be chosen. In this simulation study, we uniformly sample points in the convex hull of $\WindowIntactParticle$ for generating point patterns $\mathcal{P}$, with an adjustable number $\NumberTessellationCells\ge 1$ of points. By varying $\NumberTessellationCells$, we can generate different tessellations with varying numbers of cells, thus resulting in different numbers of generated fragments, where a larger number of points leads to a larger number of fragments, while a smaller $\NumberTessellationCells$ produces fewer fragments. For details on simulating random point patterns (also referred to as point processes), see e.g.~\cite{Chiu2013}.

Recall that we denoted the set of voxels associated with the non-broken particle by $\SetMaskImageIntact\subset \WindowIntactParticle$, see Section~\ref{Sec:Particle-based segmentation}. In order to generate virtual fragments within the non-broken particle we need to find all cells of $\Tessellation$, which overlap with $\SetMaskImageIntact$, i.e., to identify those $C\in\Tessellation$ for which $C\cap \SetMaskImageIntact\neq\emptyset$ holds. In this way, we obtain a tessellation $\Tessellation^{\prime}=\{C^{\prime}_i\}_{i=1}^{\NumberFragmentsGenerated}$ of $\SetMaskImageIntact$, consisting of $\NumberFragmentsGenerated\geq 1$ non overlapping and non empty subsets $C^{\prime}_1,\ldots,C^{\prime}_{m^\prime}$ of $\SetMaskImageIntact$ such that $\bigcup_{i=1}^{m^\prime}C^{\prime}_i=\SetMaskImageIntact$. Then, for each $i\in\{1,2,\dots,\NumberFragmentsGenerated\}$, the $i$-th virtually generated fragment corresponds to $C^{\prime}_i$.
%Then, we obtain $\NumberFragmentsGenerated>0$ generated fragments $\FragmentGenerated_1,\dots,\FragmentGenerated_{\NumberFragmentsGenerated}\subset\WindowIntactParticle$, where $\FragmentGenerated_i=\{x\in\WindowIntactParticle: x\in C^{\prime}_i\}$.
Furthermore, for each fragment $\FragmentGenerated\in\SetFragmentsGenerated$ we get the fragment image $\ImageFragmentGenerated:\WindowFragmentGenerated\rightarrow\{0,1,\dots,65565\}$ by cropping, i.e.,
\begin{align}
  \ImageFragmentGenerated(x) = \begin{cases}
       \ImageIntactParticle(x),  \qquad &\mbox{if $x \in \FragmentGenerated$,}\\
       0, \qquad &\mbox{otherwise},
   \end{cases} 
\end{align}
for each $x\in\WindowFragmentGenerated$, where $\WindowFragmentGenerated \subset \Z^3$ denotes the smallest cuboidal sampling window which contains the fragment $\FragmentGenerated$. Thus, the mapping $\ImageFragmentGenerated$, assigning each voxel of $x\in\FragmentGenerated$ the corresponding grayscale value of $\ImageIntactParticle$, can be considered to be a cutout of $\ImageIntactParticle$ representing the fragment $\FragmentGenerated$. Besides this, we consider the mask $\MaskFragmentGenerated:
\WindowFragmentGenerated\rightarrow\{0,1\}$ of $\FragmentGenerated$, where $\MaskFragmentGenerated(x)=1$ for each $x\in\FragmentGenerated$, and $\MaskFragmentGenerated(x)=0$ otherwise. 

Additionally, we consider the image $S^\prime:\WindowIntactParticle\rightarrow\{0,1,\dots,\NumberFragmentsGenerated\}$, which assigns each voxel to a generated fragment or background, i.e.,
\begin{align}\label{Eq:Labeled ground truth image}
    \GroundTruth(x) = \begin{cases}
        i, \qquad &\mbox{if  $x\in C^\prime_i$ for some  $i\in\{1,2,\dots,\NumberFragmentsGenerated\}$},\\
        0,\qquad & \text{otherwise},
    \end{cases}
\end{align}
for each $x\in\WindowIntactParticle$. This image serves as the ground truth for evaluating the reassembling algorithm.The workflow of generating fragments by means of tessellations is sketched in Figure~\ref{Fig:Schematic illustration of virtual fragment generation}.

To account for the possible rotations of fragments during fracture, we select a rotation angle $\theta\in[-\frac{\pi}{6},\frac{\pi}{6}]^3$ at random for each fragment $\FragmentGenerated\in\SetFragmentsGenerated$ . Subsequently, the images $\ImageFragmentGenerated$ and $\MaskFragmentGenerated$ are rotated by means of the rotation function $\RotationFunction:\R^3\to\R^3$, as described in Section~\ref{Sec:Mask-based image registration}. The choice of a random rotation within the range of $[-\frac{1}{6}\pi,\frac{1}{6}\pi]^3$ was intentional, as component-wise larger rotation angles led to greatly extended computation times for achieving satisfactory reassembling results. Furthermore, for the experimentally acquired datasets of ore and slag, described in Section~\ref{Sec:Material and sample preparation},  only minor rotations of fragments are typically encountered.

%%%%%%%%%%%%%%%%%%%%%%%%%%%%%%%%%%%%%%%%%%%%%%%%%%%%%%%%%%%%%%%%%%%%%%%%%%%%%%%%%%%%

\subsection{Evaluation of reassembling algorithm for virtually generated fragments}\label{Sec:Evaluation step algorithm for virtual generated fragments}

To evaluate the reassembling algorithm 
introduced in Section~\ref{Sec:Virtual reassembling algorithm}, we apply this algorithm to the virtually generated fragments as described in Section~\ref{Sec:Virtual generation of fragments}. Therefore, we consider two performance measures to evaluated the goodness of fit of the resulting reassembled particles, by comparing the image $\ImageReassemblingResultLabeled$ given in Eq.~(\ref{Eq:Update Reassembling result labeled}) with the ground truth image $\GroundTruth$ given in Eq.~(\ref{Eq:Labeled ground truth image}).
Namely, the fraction $\lambda\in[0,1]$ of correctly assigned voxels, which is given by
\begin{align}\label{Eq:Correctly assigned voxels}
    \lambda =    \frac{\lvert\{x\in\WindowIntactParticle: \GroundTruth(x) = \ImageReassemblingResultLabeled(x)\}\rvert}{\lvert\{x\in\WindowIntactParticle: \GroundTruth(x) > 0\}\rvert},
\end{align}
and the fraction $\lambda^{\mathrm{fracture}}\in[0,1]$ of correctly assigned voxels located on the fracture surface, defined as
\begin{align}\label{Eq:Correctly assigned fracture voxels}
    \lambda^{\mathrm{fracture}} =    \frac{\lvert\{x\in\WindowIntactParticle: \ImageFractureSurface(x)=\ImageFractureSurfaceReassembled(x)\}\rvert}{\lvert\{x\in\WindowIntactParticle:  \ImageFractureSurface(x)=1\}\rvert},
\end{align}
where the images $\ImageFractureSurface:\WindowIntactParticle\rightarrow\{0, 1\}$ and $\ImageFractureSurfaceReassembled:\WindowIntactParticle\rightarrow\{0, 1\}$ describe the fracture surfaces observed in $\GroundTruth$ and $\ImageReassemblingResultLabeled$, respectively. More details on the definition and computation of $\ImageFractureSurface$ and $\ImageFractureSurfaceReassembled$ are given in Section~\ref{Sec:Voxelwise determination of fracture surfaces}.

To evaluate the reassembling algorithm, we generated $20$ different sets of fragments as described in Section~\ref{Sec:Virtual generation of fragments}, where the number of fragments in these sets varies between $2$ and $13$. The performance measures provided in Eqs.~(\ref{Eq:Correctly assigned voxels}) and~(\ref{Eq:Correctly assigned fracture voxels}) were computed for each of the 20 sets of fragments, where we obtain the averaged values $\lambda=0.96$ and $\lambda^{\mathrm{fracture}}=0.69$. Note that the mean value of $\lambda^{\mathrm{fracture}}$ is smaller than that of $\lambda$. This discrepancy results from the fact that even a slight misalignment or rotation of a reassembled fragment compared to its original position or orientation leads to a relatively large mismatch between the fracture surfaces observed in $\ImageFractureSurface$ and $I_R^\mathrm{fracture}$, which leads to smaller values of $\lambda^{\mathrm{fracture}}$. Nevertheless, the original positions of the fractures are well captured by the reassembling algorithm, as the mean value of $\lambda$ is close to $1$. These results indicate that the reassembly algorithm is capable of accurately reassembling the simulated fragments.

%%%%%%%%%%%%%%%%%%%%%%%%%%%%%%%%%%%%%%%%%%%%%%%%%%%%%%%%%%%%%%%%%%%%%%%%%%%%%%%%%%%%
%%%%%%%%%%%%%%%%%%%%%%%%%%%%%%%%%%%%%%%%%%%%%%%%%%%%%%%%%%%%%%%%%%%%%%%%%%%%%%%%%%%%

\section{Quantitative analysis of fracture mechanisms}\label{Sec:Fracture characterization}

As already mentioned above, a deeper understanding of whether fractures during fragmentation of particles occur along grain boundaries (intergranular fractures), randomly (transgranular fractures) or as a superposition of both fracture mechanisms is important for the liberation of minerals. To characterize fracture mechanisms locally, we need to analyze the fracture surfaces between adjacent fragments. Therefore, in Section~\ref{Sec:Voxelwise determination of fracture surfaces} we explain how fracture surfaces can be determined voxelvise from the output of the virtual reassembling algorithm presented in Section~\ref{Sec:Virtual reassembling algorithm}. 
Then, in Section~\ref{Sec:Descriptors of fracture surfaces}, two different descriptors of fracture surfaces are considered, which characterize global and local features of fracture mechanisms. Finally, in Section~\ref{Sec:Analysis of fracture mechanisms in ore and slag}, these tools are applied in order to analyze the fracture mechanisms of two (ore and slag) composite particles, which consist of three mineral phases as described in Section~\ref{Sec:Phase-based segmentation}.

 %This enables us to voxelwise compare opposite regions at fracture surfaces. Since the reassembling algorithm proposed in Section~\ref{Sec:Virtual reassembling algorithm} provides sufficient results when reassembling generated fragments, see Section~\ref{Sec:Simulation study}, we apply this algorithm to real particles. 
%In this paper, we consider two composite particles, an ore particle and a slag particle, where each particle consists of three mineral phases as described in Section~\ref{Sec:Phase-based segmentation}. The output of the reassembly algorithm gives a deep insight into the fracture mechanisms for each particle by allowing voxelwise determination of fracture surfaces (see Section~\ref{Sec:Voxelwise determination of fracture surfaces}). In addition, different descriptors are defined in Section~\ref{Sec:Descriptors of fracture surfaces}, which facilitate the global and local determination of fracture mechanisms. These are then computed for both composite particles in Section~\ref{Sec:Analysis of fracture mechanisms in ore and slag}.

%%%%%%%%%%%%%%%%%%%%%%%%%%%%%%%%%%%%%%%%%%%%%%%%%%%%%%%%%%%%%%%%%%%%%%%%%%%%%%%%%%%%

\subsection{Voxelwise determination of fracture surfaces}\label{Sec:Voxelwise determination of fracture surfaces}

\begin{figure}[ht!]
     \centering
     \begin{subfigure}[]{0.32\textwidth}
         \centering
         \includegraphics[width=\textwidth]{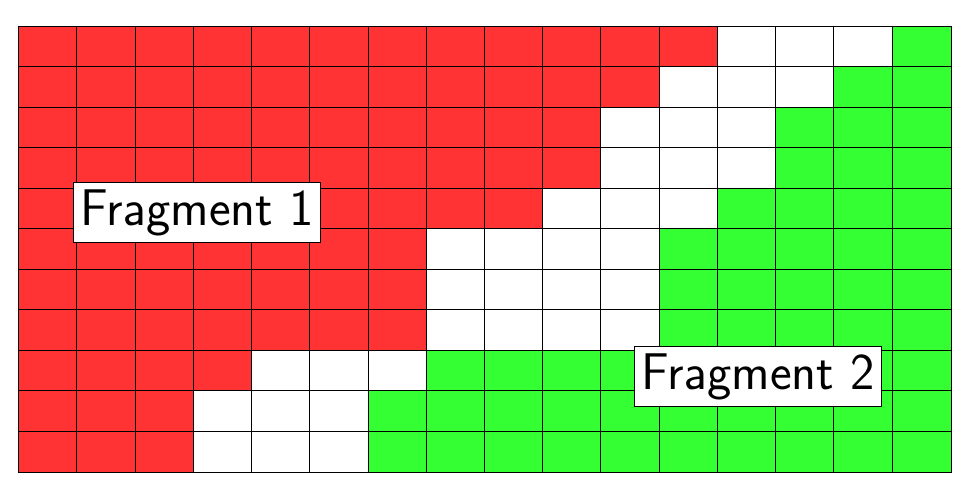}
             \subcaption{~}
     \end{subfigure}
     \begin{subfigure}[]{0.32\textwidth}
         \centering
         \includegraphics[width=\textwidth]{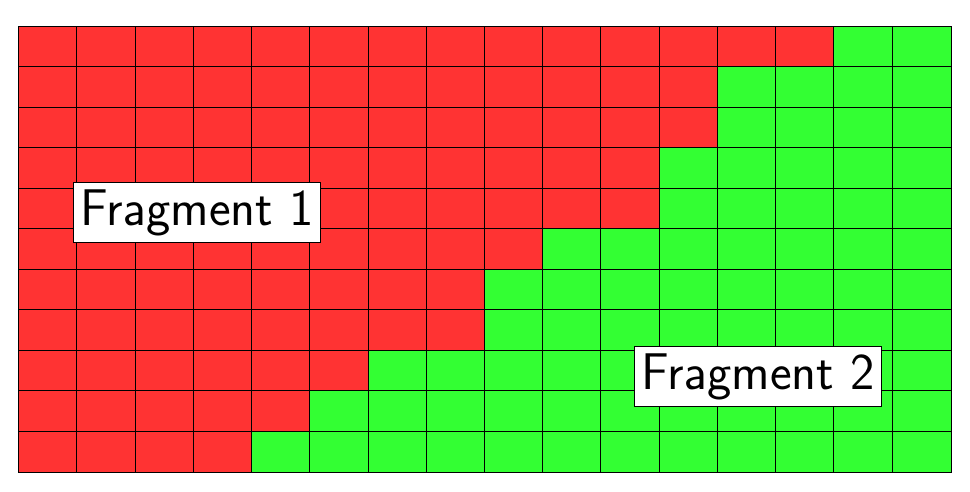}
             \subcaption{~}
     \end{subfigure}
     \begin{subfigure}[]{0.32\textwidth}
         \centering
         \includegraphics[width=\textwidth]{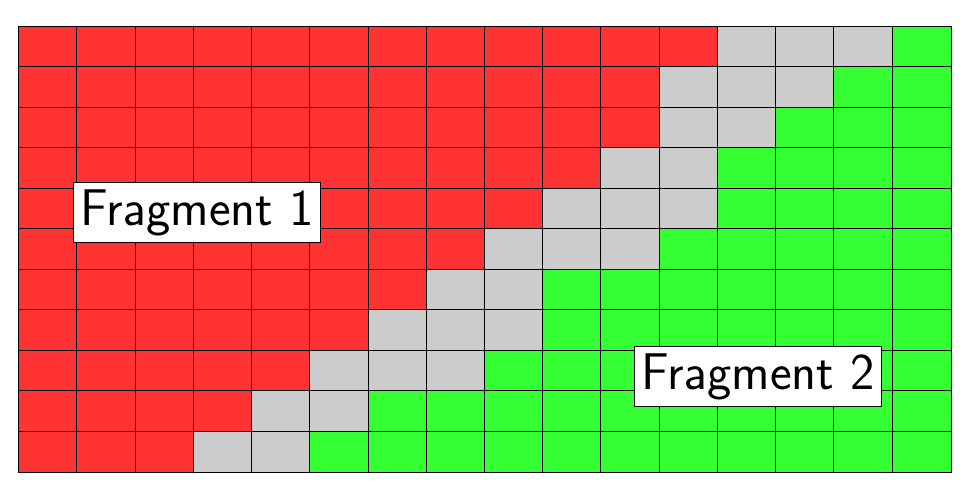}
             \subcaption{~}
     \end{subfigure}  
        \caption{Determination of fracture surfaces. (a) Gaps between two differently labeled regions corresponding to distinct fragments (colored in red and green) are identified by white voxels,  and (b)  filled using a region-growing approach. (c) The fracture surfaces are then determined by identifying all boundary voxels (colored in gray) of adjacent (differently labeled) fragments.}\label{Fig:Extraction of fracture surfaces}
\end{figure}

Recall that the output of the reassembling algorithm considered in this paper is an image $\ImageReassemblingResultLabeled:\WindowIntactParticle\rightarrow\{0,1,\dots,\NumberFragments\}$ of the virtually reassembled particle, where each fragment is associated with a label $i\in\{1,\dots,\NumberFragments\}$, as described in Section~\ref{Sec:Final output of the reassembling algorithm}. In this context, it is important to note that gaps between labeled fragments may occur in $\ImageReassemblingResultLabeled$. The number and sizes of these gaps depend on the quality of the reassembling and the discretization effects that arise when rotating discrete images during reassembling. These gaps prevent the determination of fracture surfaces directly from the reassembled fragments and need to be filled first. Therefore, we define the image $\ImageGaps:\WindowIntactParticle\rightarrow\{0,1\}$, which assigns each voxel of $\WindowIntactParticle$ either to 1 or 0, provided that the voxel belongs to a gap or not. i.e.,
\begin{align}\label{Eq:GapImage}
    \ImageGaps(x) = \begin{cases}
        1,\qquad & \text{if } \ImageIntactParticle(x)>0 \text{ and } \ImageReassemblingResultLabeled(x)=0,\\
        0,\qquad & \text{otherwise},
        \end{cases}
\end{align}
for each $x\in\WindowIntactParticle$. 
To extract the fracture surfaces in $\ImageReassemblingResultLabeled$, we first utilize a region-growing approach based on the Euclidean distance transform~\cite{Pratt2007} in order to fill the gaps resulting from slight errors of the reassembling algorithm, see Figure~\ref{Fig:Extraction of fracture surfaces}.  
More precisely, for each pair $i,j \in \{1,2,\dots,\NumberFragments\}$ with $i\neq j$, we compute the Euclidean distances of voxels in the region with label $i$ to voxels in the region with label $j$. Each voxel $x\in\WindowIntactParticle$ with $\ImageGaps(x) = 1$ is assigned to the labeled region to which it has the smallest Euclidean distance. If $x$ has the same (smallest) Euclidean distance to more than one labeled regions, it is assigned to one of these labeled regions at random.

After filling the gaps resulting from reassembling, all boundary voxels between adjacent fragments are identified. Hereby, we consider any voxel as a boundary voxel, which is not completely surrounded by voxels corresponding to the same fragment with respect to the $6$-neighborhood, where the $6$-neighborhood of a voxel $x\in\WindowIntactParticle$ is given by the set $M_{x} = \{y\in\WindowIntactParticle:\lVert x-y\rVert\leq\sqrt{2}\}$. The boundary voxels defined in this way correspond to the fracture surfaces of adjacent reassembled fragments. Thus, the fracture surfaces observed in $\ImageReassemblingResultLabeled$ are given by the binary image $\ImageFractureSurfaceReassembled:\WindowIntactParticle\rightarrow\{0, 1\}$, which assigns each voxel of $\WindowIntactParticle$ to the boundary between adjacent fragments, or the background, i.e.,
\begin{align}\label{Eq:FractureSurfaces}
    \ImageFractureSurfaceReassembled(x) = \begin{cases}
        1,\qquad & \text{if } x \text{ belongs to the boundary between adjacent reassembled fragments},\\
        0,\qquad & \text{otherwise},
        \end{cases}
\end{align}
for each $x\in\WindowIntactParticle$. 
%Analogously, in the context of the simulation study, we define the binary image $\ImageFractureSurface:\WindowIntactParticle\rightarrow\{0, 1\}$ containing the fracture surfaces observed in the ground truth image $\GroundTruth$.

% The voxels located on the fracture surfaces correspond to the boundary voxels, i.e., the fracture surfaces are given by the image $\ImageFractureSurface$, which is defined in Eq.~(\ref{Eq:FractureSurfaces}). 

%%%%%%%%%%%%%%%%%%%%%%%%%%%%%%%%%%%%%%%%%%%%%%%%%%%%%%%%%%%%%%%%%%%%%%%%%%%%%%%%%%%%

\subsection{Descriptors of fracture surfaces}\label{Sec:Descriptors of fracture surfaces}

We explore two different descriptors that characterize fracture surfaces and thus enable us to analyze whether fractures occur typically intergranular, transgranular or as a superposition of both fracture types. 
%We elaborate on the computation of material phase-based descriptors that characterize both the global and local fracture behavior.
First, we investigate the global behavior of fracture mechanisms, considering the set $\FractureVoxels$ of voxels which are located at fracture surfaces, i.e., 
\begin{align}\label{Eq:FractureVoxels}
    \FractureVoxels_{\rm R}= \{x\in\WindowIntactParticle:\ImageFractureSurface_{\rm R}(x)=1\},
\end{align}
where $\ImageFractureSurface_{\rm R}$ is given in Eq.~(\ref{Eq:FractureSurfaces}). In particular, we investigate the global fracture behavior of each material phase $i\in\{1,\dots,\NumberPhases\}$ observed in images of non-broken particles as described in Section~\ref{Sec:Phase-based segmentation}. More precisely, for each $i\in\{1,\dots,\NumberPhases\}$, the fraction $\VolumeFractionsFractureSurfaces_i\in[0,1]$ of voxels at fracture surfaces of the $i$-th phase is considered, where
\begin{align}
    \VolumeFractionsFractureSurfaces_i = \frac{\lvert\{x\in\FractureVoxels_{\rm R}:\ImagePhases(x)=i\}\rvert}{\lvert\FractureVoxels_{\rm R}\rvert}. 
\end{align}

Furthermore, we investigate the local (voxelwise) behavior of fracture mechanisms,
%by computing a descriptor for each fracture voxel. 
considering the local entropy $E(x)\in[0,\infty)$ for each $x\in\FractureVoxels_{\rm R}$, which is defined as follows. For each $x\in\FractureVoxels_{\rm R}$ and for each $i\in\{1,\dots,\NumberPhases\}$, we consider the relative frequency $p_{x}^{(i)}\in[0,1]$ of neighbors of $x$ belonging to the $i$-th material phase, where
\begin{align}\label{Eq:Classprobabilities}
    p_{x}^{(i)} = \frac{1}{6} \lvert\{y\in M_{x}:\ImagePhases(y)=i\}\rvert.
\end{align}
%where $M_{x_i}$ is the set of voxels which are in the $6$-neighborhood of $x_i$. The image $\ImagePhases$ is given in Eq.~(\ref{Eq:ImagePhases}). 
Then, the local entropy $E(x)$ of $x\in\FractureVoxels_{\rm R}$ is given by
\begin{align}\label{Eq:Entropy}
    E(x) = \sum_{i=1}^{\NumberPhases} p_{x}^{(i)}\log_2(p_{x}^{(i)}),
\end{align}
where $\log_2$ denotes the logarithm to the basis $2$ (i.e., $s=2^{\log_2(s)}$ for all $s>0$). 

Note that $E(x)$ can be interpreted as degree of disorder of the mineral phases in the local neighborhood of 
$x\in\FractureVoxels_{\rm R}$. In other words, a fracture voxel has a large value $E(x)$ if many neighboring voxels are assigned to different mineral phases. Vice versa, if $E(x)=0$, then all neighboring voxels of 
$x\in\FractureVoxels_{\rm R}$ are assigned to the same mineral phase. 
Therefore, a voxel
$x\in\FractureVoxels_{\rm R}$ with $E(x)=0$
indicates an intergranular fracture, whereas transgranular fractures are typically indicated by local entropy values larger than zero. In particular, for each $i\in\{1,\dots,\NumberPhases\}$, the transgranular fraction $\RatioTransgranular_i\in[0,1]$ of the $i$-th mineral phase is given 
 by\begin{align}
    \RatioTransgranular_i = \frac{\lvert\{x\in\FractureVoxels_{\rm R}:E(x)>0, \ImagePhases(x)=i\}\rvert}{\lvert\{x\in\FractureVoxels_{\rm R}:\ImagePhases(x)=i\}\rvert},
\end{align} 
whereas the intergranular fraction $\RatioIntergranular_i\in[0,1]$ of the $i$-th phase is given by 
\begin{align}
    \RatioIntergranular_i = 1-\RatioTransgranular_i.
\end{align}
%Consequently, we compute for each mineral phase the intergranular and transgranular fraction, see Table~\ref{Tab:Fracture specific characteristics}, where a large fraction indicates whether a mineral phase tends to break intergranularly or transgranularly.

%%%%%%%%%%%%%%%%%%%%%%%%%%%%%%%%%%%%%%%%%%%%%%%%%%%%%%%%%%%%%%%%%%%%%%%%%%%%%%%%%%%%

\subsection{Numerical results}\label{Sec:Analysis of fracture mechanisms in ore and slag}

In Table~\ref{Tab:Fracture specific characteristics},
 numerical values of the fraction $\VolumeFractionsFractureSurfaces_i$ of voxels at fracture surfaces of the $i$-th phase, as well as of its transgranular and intergranular fractions $\RatioTransgranular_i$ and $\RatioTransgranular_i$, 
are given which we obtained for the three phases of the ore and slag particles described in Section~\ref{Sec:Phase-based segmentation}.

Comparing the values of 
$\VolumeFractionsFractureSurfaces_i$ with those of the corresponding global 
volume fraction $\rho_i$, i.e., the fraction of all voxels associated with the $i$-th phase, we can observe slight discrepancies between these values,
see Table~\ref{Tab:Phase-specific fracture share}.
In particular, for
 the ore particle, the fraction 
$\VolumeFractionsFractureSurfaces_1$ of voxels at fracture surfaces
 of phase $P_1$ is smaller than the volume fraction $\rho_1$ of the entire particle, whereas the opposite behavior is observed for the other two phases $P_2$ and $P_3$ of the ore particle. 
 Similar observations are made for the slag particle, which indicates that for both particles, fractures occur more densely in phases $P_2$ and $P_3$, compared to phase $P_1$.

Given that in the ore particle, $P_1$ constitutes the phase with the largest volume fraction ($\rho_1=0.96$), fractures predominantly occur within this phase, i.e., fractures in phase $P_1$ of the ore particle occur predominantly transgranular.
%Conversely, the volume fraction of phase $P_1$ is smaller than the volume fraction of this phase in the particle volume, indicating a higher likelihood of fractures occurring in the other two phases. Additionally, it is evident that fractures in phase $P_1$ predominantly occur transgranular, as the fracture must traverse this phase before reaching other phases.
In contrast, regarding the other two phases, there is no clear indication of fractures occurring either intergranular or transgranular, i.e., it seems that both $P_2$ and $P_3$ exhibit a mixture of both fracture mechanisms.

 With respect to latter aspect, the fracture behavior of the slag particle is different.
 Furthermore, note that there is a more pronounced balancing between the volume fractions of the three mineral phases of the slag particle, i.e., there is not only one predominant phase, see
 Table~\ref{Tab:Phase-specific fracture share}.
 Besides this, the slag particle contains air bubbles in phase $P_3$ and fracture surfaces always occur transgranular in this phase.
 However, for phases $P_1$ and $P_2$ of the slag particle, there seem to be no dominant transgranular nor integranular fracture mechanisms, as the transgranular and integranular fractions 
 $\RatioTransgranular_2$, $ \RatioIntergranular_2$ as well as
 $\RatioTransgranular_3$, $ \RatioIntergranular_3$
 are significantly larger than zero, see Table~\ref{Tab:Fracture specific characteristics}.

\begin{table}[ht!]
\caption{Numerical values obtained for the descriptiors of fracture surfaces of ore and slag particles
%the fraction $\VolumeFractionsFractureSurfaces_i$ of voxels at fracture surfaces of the $i$-th phase, as well as for the transgranular fraction $\RatioTransgranular_i$ and the intergranular fraction $\RatioTransgranular_i$
($i=1,2,3)$.
\label{Tab:Fracture specific characteristics}}  
\newcolumntype{D}{>{\centering}p{0.3\textwidth}}
		\begin{tabularx}{\textwidth}{DCCC}
			\toprule
		ore	&phase $P_1$ & phase $P_2$&phase $P_3$\\
   \midrule
			$\VolumeFractionsFractureSurfaces_i$ &0.91&0.08&0.002\\
            $\RatioTransgranular_i$&0.05&0.36&0.57\\
            $\RatioIntergranular_i$ &0.95&0.64&0.43\\
			%\bottomrule
		%\end{tabularx}
		%\begin{tabularx}
  %{\textwidth}{DCCC}
			%\toprule
   \bottomrule
			slag &phase $P_1$ & phase $P_2$&phase $P_3$\\
     \midrule
			$\VolumeFractionsFractureSurfaces_i$ &0.70 & 0.24 & 0.06\\
            $\RatioTransgranular_i$ &0.37&0.68&1.00\\
            $\RatioIntergranular_i$ &0.63&0.32&0.00\\
			\bottomrule
		\end{tabularx}
\end{table}

%%%%%%%%%%%%%%%%%%%%%%%%%%%%%%%%%%%%%%%%%%%%%%%%%%%%%%%%%%%%%%%%%%%%%%%%%%%%%%%%%%%%
%%%%%%%%%%%%%%%%%%%%%%%%%%%%%%%%%%%%%%%%%%%%%%%%%%%%%%%%%%%%%%%%%%%%%%%%%%%%%%%%%%%%

\section{Conclusion}\label{Sec:Conclusion}

We introduced a novel method for characterizing fracture mechanisms in composite materials by means of 3D CT image
data. In mineral liberation, the understanding of these mechanisms is
crucial, particularly whether fractures occur along the boundaries of mineral phases (intergranular fracture) and/or within mineral
phases (transgranular fracture). Conventional techniques for analyzing fracture mechanisms are focused on globally comparing
the surface exposure of mineral phases extracted from image measurements before and after fracture. Instead, we presented a virtual
reassembling algorithm based on image registration techniques, which is applied to 3D data of two different composite particles before and
after fracture in order to determine and characterize the individual fracture surfaces. This enabled us to conduct a local quantitative
analysis of fracture mechanisms by voxelwise comparing adjacent regions at fracture surfaces. The algorithm has been validated through a simulation study and subsequently applied to image data of two composite particles: a natural occurring ore and a slag from a geometallurgical recycling process. 

Since the characterization of the fracture mechanisms cannot be performed directly on image data obtained from CT
measurements, some image preprocessing was required, where we binarized the CT image data of the ore and slag particles before and after fracture, followed by a phase-based segmentation step to extract mineral phases from the non-broken particles. However, since different mineral phases may have similar X-ray attenuation coefficients leading to similar grayscale values \cite{Englisch2023}, it can happen that the segmentation of mineral phases, merely based on CT image data, is difficult or even impossible. Therefore, in a forthcoming paper, we will apply our method for characterizing fracture mechanisms to an extended set of image data, combining 3D CT measurements of composite particles with 2D SEM-EDS data acquired by means of the mineral liberation analyzer (MLA) \cite{Furat2018, Furat2024}.

%%%%%%%%%%%%%%%%%%%%%%%%%%%%%%%%%%%%%%%%%%
%%%%%%%%%%%%%%%%%%%%%%%%%%%%%%%%%%%%%%%%%%

\vspace{1cm}
%\newpage

%%%%%%%%%%%%%%%%%%%%%%%%%%%%%%%%%%%%%%%%%%
\authorcontributions{
Conceptualization, U.P. and V.S.; 
data curation, T.V.; 
formal analysis, T.W., T.V. and O.F.; 
funding acquisition, U.P. and V.S.;
investigation, T.W., T.V. and O.F.; 
methodology, T.W., T.V. and O.F.; 
project administration, U.P. and V.S.;
resources, T.W., T.V. and O.F.; 
software, T.W.; 
supervision, U.P. and V.S.; 
visualization, T.W.; 
writing - original draft, T.W., T.V. and O.F.;  
writing - review \& editing, T.W., T.V., O.F., U.P. and V.S.;
All authors have read and agreed to the published version of the manuscript.
}

\funding{This research is partially funded by the German Research Foundation (DFG) via the research projects PE 1160/32-1 and SCHM 997/45-1 within the priority programs SPP 2315 “Engineered artificial minerals (EnAM)---A geo-metallurgical tool to recycle critical elements from waste streams”. 
}

\dataavailability{The datasets generated and/or analyzed during the current study are available from the corresponding author on reasonable request.}

\conflictsofinterest{The authors declare no conflict of interest and the funders had no role in the design of the study; in the collection, analyses, or interpretation of data; in the writing of the manuscript; or in the decision to publish the~results.}

%%%%%%%%%%%%%%%%%%%%%%%%%%%%%%%%%%%%%%%%%%
\begin{adjustwidth}{-\extralength}{0cm}

\reftitle{References}

\begin{align}\label{def.ssm}
        \SSM_{\ImageFixed,\ImageFixedAdjusted}(x)=
        \begin{cases}
            \displaystyle\frac{(2 \mu_{\ImageFixed}(x)\mu_{\ImageFixedAdjusted}(x)+ c_1)(2\sigma_{\ImageFixed,\ImageFixedAdjusted}(x)+c_2)}{(\mu_{\ImageFixed}^2(x) + \mu_{ \ImageFixedAdjusted}^2(x)+ c_1)(\sigma_{\ImageFixed}^2(x) + \sigma^2_{ \ImageFixedAdjusted}(x)+c_2)},\qquad &\text{if} (\mu_{\ImageFixed}^2(x) + \mu_{ \ImageFixedAdjusted}^2(x)+ c_1)(\sigma_{\ImageFixed}^2(x) + \sigma^2_{ \ImageFixedAdjusted}(x)+c_2) \neq 0,\\
            0, \qquad&\text{otherwise,} 
        \end{cases}
    \end{align}
 where $\mu_{\ImageFixed}(x)\in\R$ is the local mean grayscale value of $\ImageFixed$ at 
$x\in\BoundingBoxtransformedMaskOptimal$ with
    \begin{align}\label{Eq:Mu}
        \mu_{\ImageFixed}(x) =  (\ImageFixed\ast \GaussianFilter)(x),
    \end{align}
and $\ast$ denotes  convolution. The function $k:K=\{-5,\dots,5\}^3\rightarrow \R$ in Eq.~(\ref{Eq:Mu} is a (truncated) Gaussian kernel~\cite{Burger2016}, which is given by 
    \begin{align}
        \GaussianFilter(x) = \frac{1}{\nu}\exp{\bigg(\frac{\lVert x \rVert}{2\sigma^2}\bigg)}
    \end{align}    
for any $x\in K$ and $\sigma\geq 0$, where the normalizing constant $\nu>0$ is chosen such that $\sum_{x\in K}k(x) = 1$. In our case, it turned out that $\sigma=1.5$ is a suitable choice. Note that the convolution of $\ImageFixed$ and $k$ considered in Eq.~(\ref{Eq:Mu}) is defined as 
\begin{align*}
    (\ImageFixed\ast \GaussianFilter)(x) = \sum_{(i,j,k)\in K}\ImageFixed(x_1-i,x_2-j,x_3-k)\; k(i,j,k),
    \end{align*}
for each $x=(x_1,x_2,x_3)\in\BoundingBoxtransformedMaskOptimal$, where we put $\ImageFixed(x_1-i,x_2-j,x_3-k)=0$ if $(x_1-i,x_2-j,x_3-k)\notin\BoundingBoxtransformedMaskOptimal$ for some $(i,j,k)\in K$. For further details regarding convolutions, we refer to~\cite{Burger2016}. The local mean grayscale value $\mu_{\ImageFixedAdjusted}(x)\in\R$ of $\ImageFixedAdjusted$ at 
$x\in\BoundingBoxtransformedMaskOptimal$ 
 is defined analogously. Furthermore, the local standard deviation $\sigma_{\ImageFixed}(x)\in[0,\infty)$ of $\ImageFixed$ at $x\in\BoundingBoxtransformedMaskOptimal$ is given by 
    \begin{align}
        \sigma_{\ImageFixed}(x) =  ((\ImageFixed\cdot\ImageFixed)\ast \GaussianFilter)(x)-\mu^2_{\ImageFixed}(x),
    \end{align}
where $\cdot$ denotes the voxelwise multiplication of two images. The local standard deviation $\sigma_{\ImageFixedAdjusted}(x)$ of $\ImageFixedAdjusted$ is defined analogously. Finally, the local covariance $\sigma_{\ImageFixed,\ImageFixedAdjusted}(x)$ is given by
    \begin{align}
        \sigma_{\ImageFixed,\ImageFixedAdjusted}(x) = ((\ImageFixed\cdot \ImageFixedAdjusted)\ast \GaussianFilter)(x)-\mu_{\ImageFixed}(x)\mu_{\ImageFixedAdjusted}(x),
    \end{align}
for each $x\in\BoundingBoxtransformedMaskOptimal$. The constants $c_1$ and $c_2$ in Eq.~(\ref{def.ssm}) are chosen as in~\cite{Wang2004}, i.e.,
    \begin{align*}
    c_1=0.01\cdot(\max_{x\in \BoundingBoxtransformedMaskOptimal} \ImageFixed(x) - \min_{x\in \BoundingBoxtransformedMaskOptimal} \ImageFixed(x))\qquad\text{and}\qquad
    c_2=7.65\cdot(\max_{x\in \BoundingBoxtransformedMaskOptimal} \ImageFixed(x) - \min_{x\in \BoundingBoxtransformedMaskOptimal} \ImageFixed(x)).
    \end{align*}

\end{appendices}

\end{adjustwidth}
\end{document}